\documentclass[aps,prd,a4paper,twocolumn,amsmath,showpacs,superscriptaddress,nofootinbib,preprintnumbers]{revtex4-1}

\usepackage{verbatim}
\usepackage[T1]{fontenc}
\usepackage[utf8]{inputenc}
\usepackage[american]{babel}
\usepackage{epsfig}
\usepackage{graphicx}
\usepackage{booktabs}
\usepackage{multirow}
\usepackage{dcolumn}
\usepackage{amsmath}
\usepackage{subcaption}
\usepackage{mathtools}
\usepackage{amsfonts}
\usepackage{amssymb}
\usepackage{epstopdf}
\usepackage{bm}
\usepackage{siunitx}
\usepackage{braket}
\usepackage{enumitem}
\usepackage{soul}
\usepackage[table]{xcolor}
\usepackage{color}
\usepackage{transparent}
\usepackage{pifont}
\definecolor{navyblue}{rgb}{0.0, 0.0, 0.5}
\definecolor{royalblue}{rgb}{0.25, 0.41, 0.88}
\definecolor{cadmiumgreen}{rgb}{0.0, 0.42, 0.24}
\definecolor{blue-violet}{rgb}{0.54, 0.17, 0.89}
\definecolor{darkviolet}{rgb}{0.58, 0.0, 0.83}
\definecolor{orange(colorwheel)}{rgb}{1.0, 0.5, 0.0}

\usepackage{hyperref}
\hypersetup{
    colorlinks=true, % false: boxed links; true: colored links
    linkcolor=royalblue, % color of internal links (change box color with linkbordercolor)
    citecolor=magenta}

\newcommand\ee{\end{equation}}
\newcommand\be{\begin{equation}}
\newcommand\eea{\end{eqnarray}}
\newcommand\bea{\begin{eqnarray}}

\newcommand\ev{\,\mathrm{eV}}

\newcommand{\lcdm}{\Lambda\mathrm{CDM}}

\newcommand\ie{{\it i.e.}~}

\newcommand\eq[1]{Eq.~\eqref{eq:#1}}

\newcommand\fig[1]{Fig.~\ref{fig:#1}}
\newcommand\tab[1]{Tab.~\ref{tab:#1}}

\renewcommand{\vec}{\bm}

\newcommand{\nnu}{N_{\rm eff}}
\newcommand{\meff}{m^{\rm eff}_{\nu,\rm sterile}}
\newcommand{\mef}{m_{\rm eff}}
\newcommand{\mnu}{\Sigma m_\nu}
\newcommand{\slcdm}{Super-$\Lambda$CDM\,}

\newcommand{\ncl}{95\% CL\,}
\newcommand{\scl}{68\% CL\,}

\usepackage{booktabs}
\usepackage{multirow}
\usepackage{dcolumn}
\usepackage{colortbl}

\definecolor{magenta(process)}{rgb}{1.0, 0.0, 0.56}

\definecolor{darkspringgreen}{rgb}{0.09, 0.45, 0.27}

\definecolor{royalblue(web)}{rgb}{0.25, 0.41, 0.88}
\begin{document}

\title{Possible impact of non-Gaussianities on cosmological constraints in neutrino physics}

\author{Matteo Forconi}
\email{matteo.forconi@roma1.infn.it}
\affiliation{Physics Department and INFN, Universit\`a di Roma ``La Sapienza'', Ple Aldo Moro 2, 00185, Rome, Italy} 

\author{Eleonora Di Valentino}
\email{e.divalentino@sheffield.ac.uk}
\affiliation{School of Mathematics and Statistics, University of Sheffield, Hounsfield Road, Sheffield S3 7RH, United Kingdom}

\author{Alessandro Melchiorri}
\email{alessandro.melchiorri@roma1.infn.it}
\affiliation{Physics Department and INFN, Universit\`a di Roma ``La Sapienza'', Ple Aldo Moro 2, 00185, Rome, Italy}

\author{Supriya Pan}
\email{supriya.maths@presiuniv.ac.in}
\affiliation{Department of Mathematics, Presidency University, 86/1 College Street, Kolkata 700073, India}
\affiliation{Institute of Systems Science, Durban University of Technology, PO Box 1334, Durban 4000, Republic of South Africa}

\date{\today}

\preprint{}
\begin{abstract}
The search for non-Gaussian signatures in the Cosmic Microwave Background (CMB) is crucial for understanding the physics of the early Universe. Given the possibility of non-Gaussian fluctuations in the CMB, a recent revision to the standard $\Lambda$-Cold Dark Matter ($\Lambda$CDM) model has been proposed, dubbed "Super-$\Lambda$CDM". This model introduces additional free parameters to account for the potential effects of a trispectrum in the primordial fluctuations. In this study, we explore  the impact of the Super-$\Lambda$CDM model on current constraints in neutrino physics. In agreement with previous research, our analysis reveals that for most of the datasets, the Super-$\Lambda$CDM parameter $A_0$ significantly deviates from zero at over a $95\%$ confidence level. We then demonstrate that this signal might influence current constraints in the neutrino sector. Specifically, we find that the current constraints on neutrino masses may be relaxed by over a factor of 2 within the Super-$\Lambda$CDM framework, thanks to the correlation present with $A_0$; relaxation persists even when we apply constraints to the trispectrum from local measurements, assuming a strong similarity between the two forms. Consequently, locking $A_0=0$ might introduce a bias, leading to overly stringent constraints on the total neutrino mass.

\end{abstract}

\maketitle

%%%%%%%%%%%%%%%%%%%%%

\section{Introduction}

Presently, the best representation of our Universe and its evolution is described by the so-called $\Lambda$-Cold Dark Matter ($\Lambda$CDM) model. In this model, the dark energy is represented by a positive cosmological constant $\Lambda$, while the dark matter is considered to be pressureless or cold. The $\Lambda$CDM model is constructed using only six cosmological parameters  and it has proven to provide a reliable description of the properties of our Universe.

Nonetheless, despite its tremendous success in explaining a range of astronomical observations, the standard $\Lambda$CDM model is not exempt from several ongoing debates and challenges. Several unresolved issues still persist within this simple paradigm.  
One of the most significant and pressing concerns is the Hubble constant ($H_0$) tension, as extensively discussed in recent literature~\cite{Knox:2019rjx,DiValentino:2020zio,Jedamzik:2020zmd,DiValentino:2021izs,Schoneberg:2021qvd,Perivolaropoulos:2021jda,Shah:2021onj,Verde:2019ivm,Abdalla:2022yfr,Kamionkowski:2022pkx}. This tension highlights a massive discrepancy between the early-time estimates of $H_0$ by the Cosmic Microwave Background (CMB), such as Planck's observations ($H_0=67.4\pm0.5$ km~s$^{-1}$Mpc$^{-1}$~\cite{Planck:2018vyg}), assuming a $\Lambda$CDM model in the background, and the late-time measurements of $H_0$ using a cosmic distance ladder, such as the SH0ES (supernovae and $H_0$ for the equation of state of dark energy) Collaboration where type Ia Supernovae are calibrated with Cepheids, leading to $H_0=73.04\pm1.04$ km~s$^{-1}$~Mpc$^{-1}$~\cite{Riess:2021jrx}.

As the struggle to solve the well-known cosmological tensions between late- and early-time observations within the standard $\Lambda$CDM model is far from being solved, the possibility of observing beyond-standard-model physics is both tantalizing and promising.  A particularly insightful approach for testing new physics involves exploring departures from Gaussianity in the primordial perturbations. The CMB anisotropies are well suited for such observations. They are generated early in the Universe's history, distinct from other processes imprinting non-Gaussianities (NG) at later times, and result from the evolution of the primordial fluctuations of the field that drives inflation. The simplest inflationary models, which are constructed starting from a single field slowly rolling down to a flat potential, imply that the evolution of the field fluctuations assumes the form of a harmonic oscillator equation, where the ground state is a Gaussian wave function. Consequently, the standard prediction of an inflationary epoch directs that the cosmic fluctuations are Gaussian on the large scale. Despite being a first-order approximation, on such models, the departures from Gaussianity are negligible~\cite{Maldacena:2002vr,Acquaviva:2002ud}. However, the possibility of observing non-negligible NG is not completely ruled out. By venturing beyond the most basic realizations of the inflationary epoch, higher-order correlations could become nonzero, providing valuable insight into nonlinear interactions. Even though current measurements of NG are within the expected prediction of the single-field slow-roll model, the current level of sensitivity cannot allow us to rule out alternative theories~\cite{Planck:2019kim}. 

The scope of this work is to extend the works of Refs.~\cite{Adhikari:2019fvb,Adhikari:2022moo}. The authors of Refs.~\cite{Adhikari:2019fvb,Adhikari:2022moo} first introduced the \slcdm model, an extension of the $\Lambda$CDM model where NG are parametrized by an additional parameter in the angular power spectrum due to super sample signal (i.e. \textit{Super}) and the inflationary paradigm is described by the quasi-single-field inflationary case. Precedent works found the existence of possible NG at \ncl that is increased at more than $3\sigma$ when the curvature of our Universe is included in the picture.

If the \slcdm model proves to be accurate, it could profoundly impact current cosmological constraints that rely on the $\Lambda$CDM assumption. Specifically, in this work, we aim to examine the potential consequences for the current cosmological constraints in the neutrino sector.  Such an inquiry is highly pertinent at this time, as direct constraints of comparable precision regarding parameters like the neutrino mass are beginning to emerge from ground-based laboratories (see e.g.~\cite{Capozzi:2021fjo}). Any potential discrepancy between laboratory findings and current cosmological constraints, which are based on the $\Lambda$CDM assumption, would undoubtedly necessitate a reevaluation of the $\Lambda$CDM framework.

The article is organized as follows: in~\autoref{model} we present the model considered in this work; in~\autoref{method} we introduce the methodology and the datasets explored; in~\autoref{results} we show the results we obtained; and finally in~\autoref{concl} we derive our conclusions.

\section{Model} \label{model}

\subsection{Quasi-single-field inflationary model}

In this work, our focus centers on the quasi-single-field inflation~\cite{Chen:2009zp,Welling:2019bib,Noumi:2012vr}, a class of models which naturally emerges when considering UV completion. These models occupy a middle ground between single-field and multifield realizations of the inflationary theory. They are distinguished by a coupling between the inflaton and massive scalar fields. This interaction has the potential to produce large NG distinguishable from the single-field inflation~\cite{Chen:2009we} when the mass is of the order of the Hubble parameter. In fact, when $m\gg \mathcal{O}(H)$, the predictions align with those of single-field inflation.

The field space can be subdivided into a flat slow-roll direction, where the massless Goldstone boson inflaton $\phi$ lies, and orthogonal directions corresponding to the massive isocurvaton modes $\sigma$. The morphology of the inflaton trajectory is crucial, as, if it is straight, we end up with the conventional slow-rolling scenario, whereas a turning trajectory is responsible for the coupling between $\phi$ and $\sigma$. The latter possibility enables the conversion of fluctuations in $\sigma$ into fluctuations in the inflaton. This mechanism allows for the preservation of the NG generated by $\sigma$, because, if not transferred, they are destined to decay exponentially. Specifically, outside the horizon, the amplitude of $\delta\sigma$ is proportional to $(-\tau)^{3/2- \nu}$, where $\nu=\sqrt{9/4-m^2/H^2}$. In the case where $m/H\sim 0$, there is no decay, leading to the realization of multifield inflation. For $0<m^2/H^2<9/4$, the modes decay with a mass-dependent rate. Moreover, if $m^2/H^2>9/4$, the exponent acquires an imaginary part and assumes the characteristic form of an underdamped oscillator, suppressing the transfer of such fluctuations. Our quasi-single-field paradigm impose that $0<\nu<3/2$. 

Since, at leading order, $\delta\phi$ corresponds to curvature perturbations $\zeta$, which remain constant after horizon exit, the conversion of fluctuations preserve NG. Consequently, quantum fluctuations of the massive field can make substantial contribution to the final curvature perturbation~\cite{Meerburg:2016zdz}.

A noteworthy feature of perturbations in this regime is the fact that the trispectrum amplitude $\tau_{NL}$ is boosted with respect to the bispectrum amplitude $f_{NL}$~\cite{Baumann:2012bc,Baumann:2011nk,Assassi:2012zq}. Let us first introduce the \textit{collapsed limit} of an N-point function as the limit where one internal momentum is smaller than all the external ones (the internal momentum is the vectorial sum of M external momenta). From the Suyama-Yamaguchi inequality~\cite{Suyama:2007bg} we know that the \textit{collapsed limit} of the four-point function is constrained from below by the amplitude in the \textit{squeezed limit} of the three-point function. More precisely, $\tau_{NL}\gtrsim (\frac65 f_{NL})^2$. This inequality is particularly sensitive to the presence of multiple sources; if only one source is considered, it reaches saturation. In the quasi-single-field scenario, the trispectrum is said to be boosted because it is possible to show that
\begin{equation}
    \tau_{NL} \sim \frac{(\frac65 f_{NL})^2}{(\frac{\rho}{H})^2}
\end{equation}
where $\rho$ represents the coupling. Therefore, for a small coupling we have the amplitude of the collapsed four-point function much greater than the amplitude of the squeezed three-point function. We shall highlight that constraints for the amplitude of the squeezed 3-point function exist~\cite{refId0} but not for $\tau_{NL}(\epsilon)$.

Finally, we want to present the explicit form of the \textit{collapsed} trispectrum in the quasi-single-field models. Considering the interchangeability between curvature and inflaton, we have~\cite{Assassi:2012zq}
\begin{multline}
    \langle \zeta_{\vec{k_1}}\zeta_{\vec{k_2}}\zeta_{\vec{k_3}}\zeta_{\vec{k_4}}\rangle'_c \xrightarrow{k_{12\rightarrow0}}\\
    4\tau_{NL}(\nu)P_\zeta(k_1)P_\zeta(k_3)P_\zeta(k_{12})\left(\frac{k_{12}}{\sqrt{k_1k_3}}\right)^{3-2\nu}\label{eq:trispectrum}
\end{multline}
with $\vec{k}_{12}=\lvert k_1+k_2\rvert$ and $\langle\cdot\rangle'=\langle\cdot\rangle(2\pi)^3\delta(\vec{k}_1+\vec{k}_2+\vec{k}_3+\vec{k}_4)$ . 

\subsection{Super-$\Lambda$CDM}
By introducing $\epsilon=3/2-\nu$ into \eq{trispectrum} and performing the shift $n_s\rightarrow n_s+\epsilon$ for $k_i$ and $n_s\rightarrow n_s-2\epsilon$ for $k_{12}$, we can derive the full-sky expression for the NG covariance for the contribution of~\eq{trispectrum} (see~\cite{Adhikari:2019fvb} and the references therein)
\begin{equation}
    \mathcal{C}_{NG}=\frac9\pi\tau_{NL}(\epsilon)C^{SW}_{L=0}(n_s-2\epsilon)C_\ell(n_s+\epsilon)C_{\ell'}(n_s+\epsilon)
    \label{eq:covNG}
\end{equation}
where $C_\ell$'s are the lensed harmonics~\cite{2012JCAP...03..011P}  and $C_L^{SW}$ is the Sachs-Wolfe angular power, defined as
\begin{equation}
    C^{SW}_{L}(n_s-2\epsilon)=\frac{4\pi A_s}{9(k_0r_\star)^a}\frac{\sqrt{\pi}\Gamma(1-\frac{a}{2})\Gamma(L+\frac{a}{2})}{4\Gamma(\frac32-\frac{a}{2})\Gamma(2+L-\frac{a}{2})}
    \label{eq:CSWL}
\end{equation}
where $a=n_s-2\epsilon-1$, $0<a<2$ and $r_\star$ is the comoving distance to the last scattering surface. 

If we compute the \textit{soft limits} by  splitting the fields into short and long modes ($\phi=\phi_L+\phi_S$ and $\sigma=\sigma_L+\sigma_S$)~\cite{Assassi:2012zq} it becomes more evident that the interaction couples long and short modes. We can take advantage of this fact and include the trispectrum contribution to the covariance matrix using the Super-sample method~\cite{Li:2014jr,Manzotti:2014wca}.

The main idea is that, even though they are not directly observable, large modes (larger than the survey scale, i.e. super-sample modes) affect the evolution of small modes (sub-sample modes). As a consequence, the power spectrum is affected. Rather than considering this impact as an additional source of noise, we treat it as an extra parameter with the same analysis pipeline. This approach allows us to examine the response of the power spectrum alongside other parameters, streamlining the data analysis process. As the effect of this parameter should be equal to a noise contribution, its mean value has to be zero, whereas its variance should be set in a way that \eq{covNG} is recovered. Therefore, in our case of study, taking under consideration the trispectrum consistency condition for super sample-signal~\cite{Li:2014sga,Takada:2013wfa}, the power spectrum can be modified as follows:
\begin{equation}
    C^{m}_\ell=C_\ell-A_0C_\ell(n_s+\epsilon)
    \label{eq:Cl-A0}
\end{equation}
where $m$ stands for the power spectrum measured in the presence of the NG whereas $C_\ell$ represents the CMB power spectrum for a realization without super-sample signal coupling. $A_0$ is our additional parameter which quantifies the contribution of the trispectrum. Hence, we need to impose the condition
\begin{equation}
    \langle A_0^2\rangle= \frac{9}\pi \tau_{NL}(\epsilon)C^{SW}_{L=0}(n_s-2\epsilon)\,.
\end{equation}

It is important to underline that the latest constraints on $\tau_{NL}$~\cite{Marzouk:2022utf} are referred to the local trispectrum for multifield inflation (\ie $\epsilon=0$). For a vanishing $\epsilon$, the Sachs-Wolfe angular power spectrum computed at $L=0$ diverges for small modes, and, therefore, it cannot be directly related to our constraints of $A_0$. 

Taking into account the framework we have outlined, we can promote the $\Lambda$CDM model to the \textbf{\slcdm} model, i.e., to the usual six cosmological parameters, we add $\epsilon$, originating from the quasi-single-field framework \eq{trispectrum}, and $A_0$ introduced within the Super-sample approach. This extension of the standard model, initially introduced in Refs.~\cite{Adhikari:2019fvb,Adhikari:2022moo}, provides a more convenient picture to study the NG. The response of the CMB power spectrum in the Super$\Lambda$CDM scenario was already addressed in~\cite{Adhikari:2019fvb}. Specifically, this scenario was investigated using the temperature power spectrum from Planck 2015 release \cite{Planck:2015bpv} also in conjunction with Pantheon type Ia supernovae~\cite{Pan-STARRS1:2017jku} and Baryon Acoustic Oscillations data \cite{Beutler:2011hx,Ross:2014qpa,Gil-Marin:2015nqa}. Additionally, in Ref.~\cite{Adhikari:2022moo}, the curvature of our Universe was included and constrained the resulting scenario using both polarization and temperature spectra from Planck 2018~\cite{Planck:2018vyg,Planck:2019nip}, Pantheon type Ia supernovae~\cite{Pan-STARRS1:2017jku} and SH0ES \cite{Riess:2020fzl}. Our purpose is to add new constraints to the parameters by using different datasets. We also want to explore possible correlations among NG and neutrino physics.

\section{Methodology and Dataset} \label{method}
When performing our data analysis, we need to compute the difference between the theoretical and observed CMB power spectra.
Therefore, to take into account the response parametrized in \eq{Cl-A0}, we have modified our theoretical code to generate the theoretical angular power spectrum
\begin{equation}
    C_\ell\rightarrow C_\ell+A_0C_\ell(n_s+\epsilon)\,.
    \label{eq:CAMBRelation}
\end{equation}

As mentioned at the end of the previous section, we studied the \slcdm model putting constraints to its eight cosmological parameters, and extensions to it. 
 Moreover, we consider the following parameters to study the potential consequences in the neutrino sector:

\begin{itemize}
   
    \item $\mathbf{N_{\rm eff}}$ is the \textit{effective} number of neutrino species in the Universe~\cite{Mangano:2001iu,Bennett:2019ewm}. If neutrino decoupling had been instantaneous, we would have $N_{\rm eff}=3$. However, due to noninstantaneous decoupling, some energy is transferred to neutrinos during $e^+e^-$ annihilation, leading to a slightly higher effective number. In fact, in the standard model we have $N_{\rm eff}\approx 3.04$~\cite{Akita:2020szl,Froustey:2020mcq,Bennett:2020zkv,Cielo:2023bqp}. Any additional relativistic particle produced before recombination can be treated as an additional contribution to this number, and even primordial gravitational waves can contribute to it~\cite{Maggiore:1999vm,Giare:2022wxq}. Therefore, observing a $\Delta N_{\rm eff}\neq 0$ (\ie deviation from the standard value) can be a hint of new physics. Conversely, smaller values suggest a lower-temperature reheating~\cite{deSalas:2015glj} than expected in the $\lcdm$ universe. Notably, as the radiation energy density $\rho_r$ is proportional to the effective number of neutrinos, different values of $N_{\rm eff}$ modify the sound horizon at recombination. In particular, larger values decrease the horizon and, consequently, require higher values of $H_0$ (and $\sigma_8$) potentially moving towards late-time $H_0$ measurements. However, we cannot take this possibility with pure optimism, as the increased value of the $\sigma_8$ parameter exacerbates tensions with large-scale structure data~\cite{DiValentino:2020vvd}.

 \item $\mathbf{\sum{m_\nu}}$ is the sum of neutrino masses (see e.g.~\cite{Lesgourgues:2006nd}). The Planck $\Lambda$CDM base model assumes a normal mass hierarchy, with the minimal mass $\sum{m_\nu}=0.06$ eV. However, it is worth noting that the case of the smallest mass splitting does not determine the value, and $\sum{m_\nu}>0.06$ eV remains a plausible possibility. On the other hand, an inverted hierarchy increases the lower bound to be  $\sum{m_\nu}>0.1$ eV; thus, a stringent upper bound can exclude the latter scenario. In general, this extension can be considered the best motivated as laboratory experiments confirm that at least two neutrinos are massive~\cite{DeSalas:2018rby,Gariazzo:2023joe}. Also, cosmological probes provide constraints on the sum of neutrino masses (see, e.g.,~\cite{Lattanzi:2017ubx,Tanseri:2022zfe,DiValentino:2023fei,DiValentino:2021hoh,RoyChoudhury:2018gay,RoyChoudhury:2019hls,Craig:2024tky}). The effect of massive neutrinos is to suppress power on scales smaller than their free-streaming scale, which can be related to the reduction of lensing potential. Therefore, due to the \textit{lensing anomaly}, caution is advised when interpreting Planck results, as they might yield an overly strong upper limit~\cite{Capozzi:2021fjo,DiValentino:2021imh,diValentino:2022njd}. Additionally, increasing the neutrino mass intensifies the cosmological tension, as it leads to lower values of $H_0$.
    
    \item $\mathbf{m^{\rm eff}_{\nu,\rm sterile}}$ is the effective mass of sterile neutrinos~\cite{Asaka:2005an,Mastrototaro:2021wzl,Asadi:2022nj,Gelmini:2019deq}.
    For example, if the sterile neutrinos were to thermalize with the same temperature as active neutrinos, we should expect $N_{\rm eff}\approx 4$. However, to maintain generality, we can equally consider an arbitrary temperature $T_s$ or a distribution proportional to the active-sterile neutrino mixing angle~\cite{Dodelson:1993je}. In this case, a relationship between the effective mass and the physical mass of sterile neutrinos can be derived through $N_{\rm eff}$. For instance, within the context of a thermally distributed scenario, we have $m_{\nu,\rm sterile}=(\Delta N_{\rm eff})^{3/4}m^{\rm thermal}_{\rm sterile}$. We can see that for small values of the effective number of relativistic species, the physical mass $m^{\rm thermal}_{\rm sterile}$ increases. Consequently, neutrinos become nonrelativistic before recombination. A limit to the physical mass is required in order to leave out the cases where sterile neutrinos can be considered a candidate for warm and cold dark matter~\cite{Planck:2013pxb}. Specifically, we set $m^{\rm thermal}_{\rm sterile}<10$ eV, as done by the Planck Collaboration~\cite{Planck:2018vyg}.
\end{itemize}

For all the different cosmological parameters, we choose flat-prior distributions, varying them uniformly in the conservative ranges listed in \tab{Priors}. Then, for each model, we perform Monte Carlo Markov chain (MCMC) analyses using the publicly available package Cobaya~\cite{Torrado:2020dgo} and computing the theory with our modified version of CAMB~\cite{Lewis:1999bs,2012JCAP} according to \eq{CAMBRelation}. We explore the posteriors of our parameter space using the MCMC sampler developed for CosmoMC~\cite{Lewis:2002ah,Lewis:2013hha} and tailored for parameter spaces with a speed hierarchy which also implements the ``fast dragging” procedure~\cite{neal2005taking}. The convergence of the chains obtained with this procedure is tested using the Gelman-Rubin criterion~\cite{Gelman:1992zz}, and we choose as a threshold for chain convergence $R-1\lesssim 0.02$.  The likelihoods we decided to use are the following:
\begin{itemize}
    \item CMB temperature and polarization power spectra from the legacy Planck release~\cite{Planck:2018vyg,Planck:2019nip} with  \textit{plik}TTTEEE+lowl+lowE, which we will call from now on, Planck.
    
    \item Lensing Planck 2018 likelihood~\cite{Planck:2018lb}, reconstructed from the measurements of the power spectrum of the lensing potential. We refer to this dataset as just Lensing.
    
    \item Baryon Acoustic Oscillations (BAO) measurements extracted from data from the 6dFGS~\cite{Beutler:2011hx}, SDSS MGS~\cite{Ross:2014qpa}, BOSS DR12~\cite{Alam:2016hwk} and eBOSS DR16~\cite{eBOSS:2020yzd} surveys. 
    We call this dataset BAO. 
    
    \item Pantheon sample which consists of $1048$ type Ia supernovae measurements spanning the redshift range $0.01<z<2.3$~\cite{Pan-STARRS1:2017jku}.

\end{itemize}

\begin{table}
	\begin{center}
		\renewcommand{\arraystretch}{1.5}
		\begin{tabular}{c@{\hspace{0. cm}}@{\hspace{1.5 cm}} c}
			\hline
			\textbf{Parameter}    & \textbf{Prior} \\
			\hline\hline
			$\Omega_{\rm b} h^2$         & $[0.005\,,\,0.1]$ \\
			$\Omega_{\rm c} h^2$     	 & $[0.001\,,\,0.99]$\\
			$100\,\theta_{\rm {MC}}$     & $[0.5\,,\,10]$ \\
			$\tau$                       & $[0.01\,,\,0.8]$\\
			$\log(10^{10}A_{\rm S})$     & $[1.61\,,\,3.91]$ \\
			$n_{\rm s}$                  & $[0.8\,,\, 1.2]$ \\
			\hline
            $A_0$                        & $[-0.6\,,\,0.6]$\\
            $\epsilon$                 & $[-1\,,0]$\\
            $N_{\rm eff}$                & $[1\,,\,5]$\\
            $m^{\rm eff}_{\nu,\rm sterile}$ [eV]        & $[0\,,\,3]$\\
            $\Sigma m_\nu$ [eV]                & $[0\,,\,2]$\\
            \hline\hline
		\end{tabular}
		\caption{List of the parameter priors. The cutoff for $\epsilon$ is due to the fact that at $\epsilon=0$ we leave the quasi-single-field model and obtain the trispectrum for a multifield inflationary model. 
  }
		\label{tab:Priors}
	\end{center}
\end{table}

%---------LCDM-----------%
\begin{table*}
	\begin{center}
		\renewcommand{\arraystretch}{1.5}
		\resizebox{0.9\textwidth}{!}{\begin{tabular}{c c c c  c  c}
  	        \hline
			\textbf{Parameter} & \textbf{Planck}  & \textbf{Planck+Lensing}  & \textbf{Planck+BAO} & \textbf{Planck+Pantheon} & \textbf{Planck+All} \\
			\hline\hline
			$\Omega_\mathrm{b} h^2$ & $0.02258\pm 0.00017$& $0.02250\pm 0.00016$& $0.02255\pm 0.00015$& $0.02259\pm 0.00017$& $0.02251\pm 0.00014$\\
			$\Omega_\mathrm{c} h^2$ & $0.1185\pm 0.0016$&$0.1187\pm 0.0015$& $0.1187\pm 0.0011$& $0.1183\pm 0.0014$& $0.11861\pm 0.00099$\\
			$100\theta_\mathrm{MC}$ & $1.04110\pm 0.00033$&$1.04105 \pm 0.00033$&$1.04107\pm 0.00029$&$1.04112\pm 0.00032$&$1.04106\pm 0.00029$\\
            $\tau_\mathrm{reio}$ & $0.0516\pm 0.0086$&$0.0512\pm 0.0085$&$0.0514\pm 0.0087$&$0.0515\pm 0.0084$&$0.0510\pm 0.0085$\\
			$\log(10^{10} A_\mathrm{s})$ & $3.160\pm 0.050$&$3.081\pm 0.027$&$3.143\pm 0.043$&$3.162\pm 0.048$&$3.081\pm 0.024$\\
            $n_\mathrm{s}$ & $0.952\pm 0.012$&$0.9586\pm 0.0076$&$0.951\pm 0.011$&$0.952\pm 0.011$&$0.9584\pm 0.0075$\\
			\hline	
            $H_0$ [km s$^{-1}$ Mpc$^{-1}$] & $68.13\pm 0.71$&$67.99\pm 0.68$&$68.01\pm 0.49$&$68.20\pm 0.65$&$68.01\pm 0.45$\\
            $\sigma_8$ & $0.849\pm 0.019$&$0.8190\pm 0.0085$&$0.843\pm 0.016$&$0.849\pm 0.018$&$0.8190^{+0.0084}_{-0.0095}$\\
            $S_8$ & $0.857\pm 0.020$&$ 0.828\pm 0.013$&$ 0.852\pm 0.017$&$ 0.856\pm 0.019$&$ 0.828\pm 0.011$\\
            $A_{\rm 0}$& $-0.116\pm 0.048$&$-0.045\pm 0.029$&$-0.101\pm 0.043$&$-0.118\pm 0.045$&$-0.046^{+0.031}_{-0.027}$\\
            $\epsilon$ &$> -0.369$&$> -0.605$&$> -0.384$&$> -0.325$&$> -0.573$\\
\hline\hline    
		\end{tabular}}
	\end{center}
	\caption{Results for \slcdm. The constraints on parameters are at $68\%$ CL, while upper bounds are at $95\%$ CL.}
	\label{tab:lcdm}
\end{table*}

\begin{figure*}[htp]
	\centering
	\includegraphics[width=0.9\textwidth]{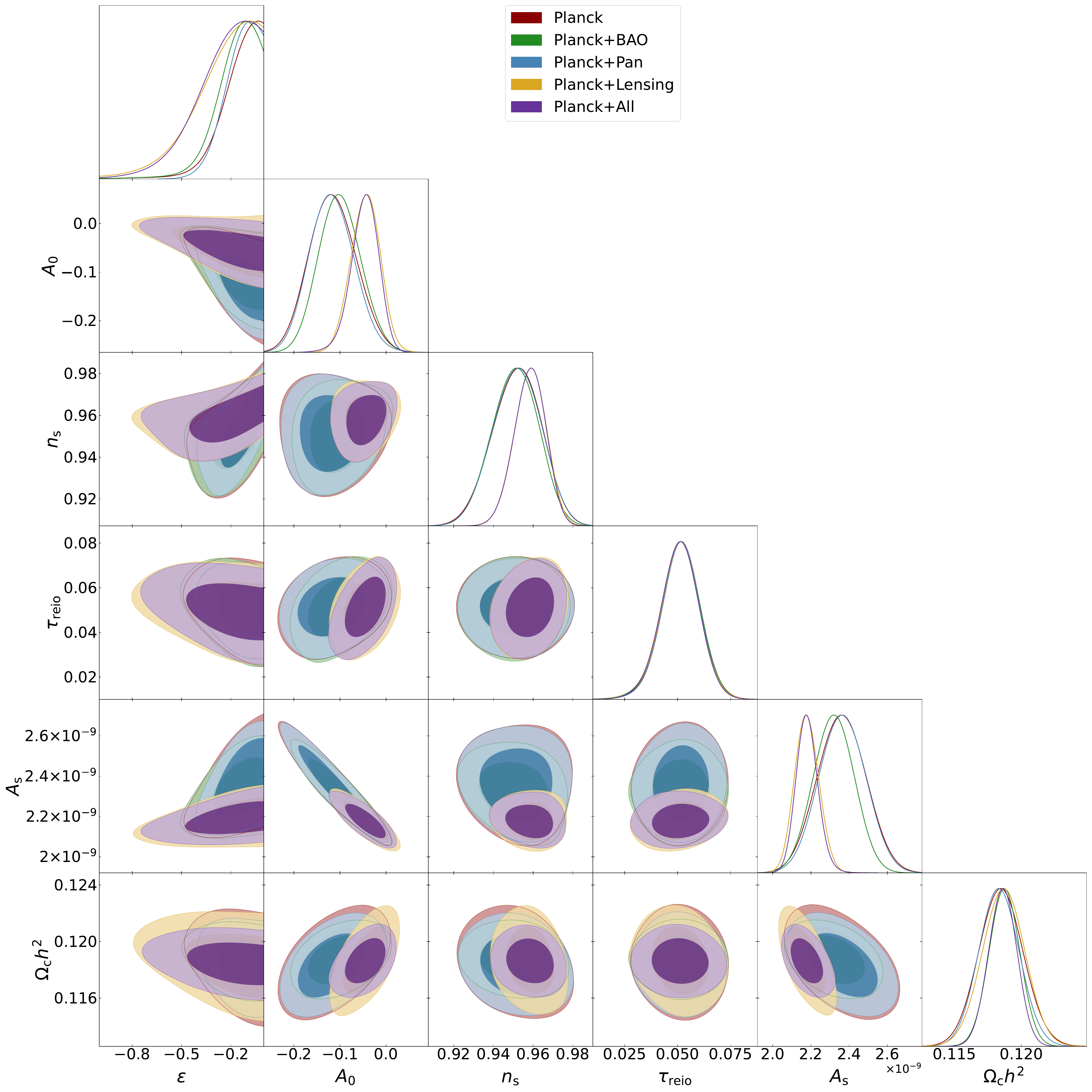}
	\caption{Marginalized 2D and 1D posterior distributions for the \slcdm.}
	\label{fig:lcdm}
\end{figure*}

%---------Neff-----------%
\begin{table*}
	\begin{center}
		\renewcommand{\arraystretch}{1.5}
		\resizebox{0.9\textwidth}{!}{\begin{tabular}{c c c c  c  c}
  	        \hline
			\textbf{Parameter} & \textbf{Planck}  & \textbf{Planck+Lensing}  & \textbf{Planck+BAO} & \textbf{Planck+Pantheon} & \textbf{Planck+All} \\
			\hline\hline
			$\Omega_{\rm b} h^2$ &$0.02248\pm 0.00026$&$0.02234\pm 0.00024$&$0.02247\pm 0.00019$ & $0.02252\pm 0.00024$ &$0.02241\pm 0.00019$\\
			$\Omega_{\rm c} h^2$ &$0.1170\pm 0.0031$&$0.1161\pm 0.0030$&$0.1166\pm 0.0030$ & $0.1171\pm 0.0031$ &$0.1163\pm 0.0029$\\
			$100\,\theta_{\rm {MC}}$ &$1.04128\pm 0.00046$&$1.04136\pm 0.00046$&$1.04133\pm 0.00046$& $1.04127\pm 0.00046$ &$1.04134\pm 0.00045$\\
			$\tau$   &$0.0512\pm 0.0087$&$0.0503\pm 0.0084$&$0.0515\pm 0.0085$& $0.0514\pm 0.0087$ &$0.0509\pm 0.0086$\\
			$\ln(10^{10}A_{\rm S})$ &$3.148\pm 0.056$&$3.068\pm 0.029$&$3.139\pm 0.043$ & $3.153\pm 0.052$ &$3.075\pm 0.024$\\
			$n_s$ &$0.945\pm 0.016$&$0.950\pm 0.012$&$0.944\pm 0.015$ & $0.0.947\pm 0.016$ &$0.952\pm 0.011$\\
			\hline	
            $H_0$ [km s$^{-1}$ Mpc$^{-1}$] &$67.3\pm 1.7$ &$66.7\pm 1.6$&$67.2\pm 1.2$&$67.6\pm 1.6$&$67.1\pm 1.1$\\
$\sigma_8$& $0.842^{+0.024}_{-0.026}$&$0.810\pm 0.013$&$0.837\pm 0.018$&$0.844\pm 0.023$&$0.812\pm 0.011$\\
$S_8$ &$ 0.855\pm 0.019$&$ 0.827\pm 0.013$&$ 0.849\pm 0.018$&$ 0.853\pm 0.020$&$ 0.824\pm 0.011$\\
$A_{\rm 0} $ &$-0.109\pm 0.050$&$ -0.040^{+0.032}_{-0.027}$&$ -0.101\pm 0.040$&$ -0.113\pm 0.046$&$ -0.045^{+0.030}_{-0.027}$\\
$\epsilon $&$> -0.420$&$> -0.692$&$ >-0.391$&$> -0.388$&$> -0.632$\\
$N_{\rm eff}$&$ 2.93\pm 0.21$&$2.86\pm 0.20$&$2.91\pm 0.18$&$2.96\pm 0.20$&$2.90\pm 0.17$\\
\hline\hline
            
		\end{tabular}}
	\end{center}
	\caption{Results for \slcdm+$N_{\rm eff}$. The constraints on parameters are at $68\%$ CL, while upper bounds are at $95\%$ CL.}
	\label{tab:nnu}
\end{table*}

\begin{figure*}[htp]
	\centering
	\includegraphics[width=0.9\textwidth]{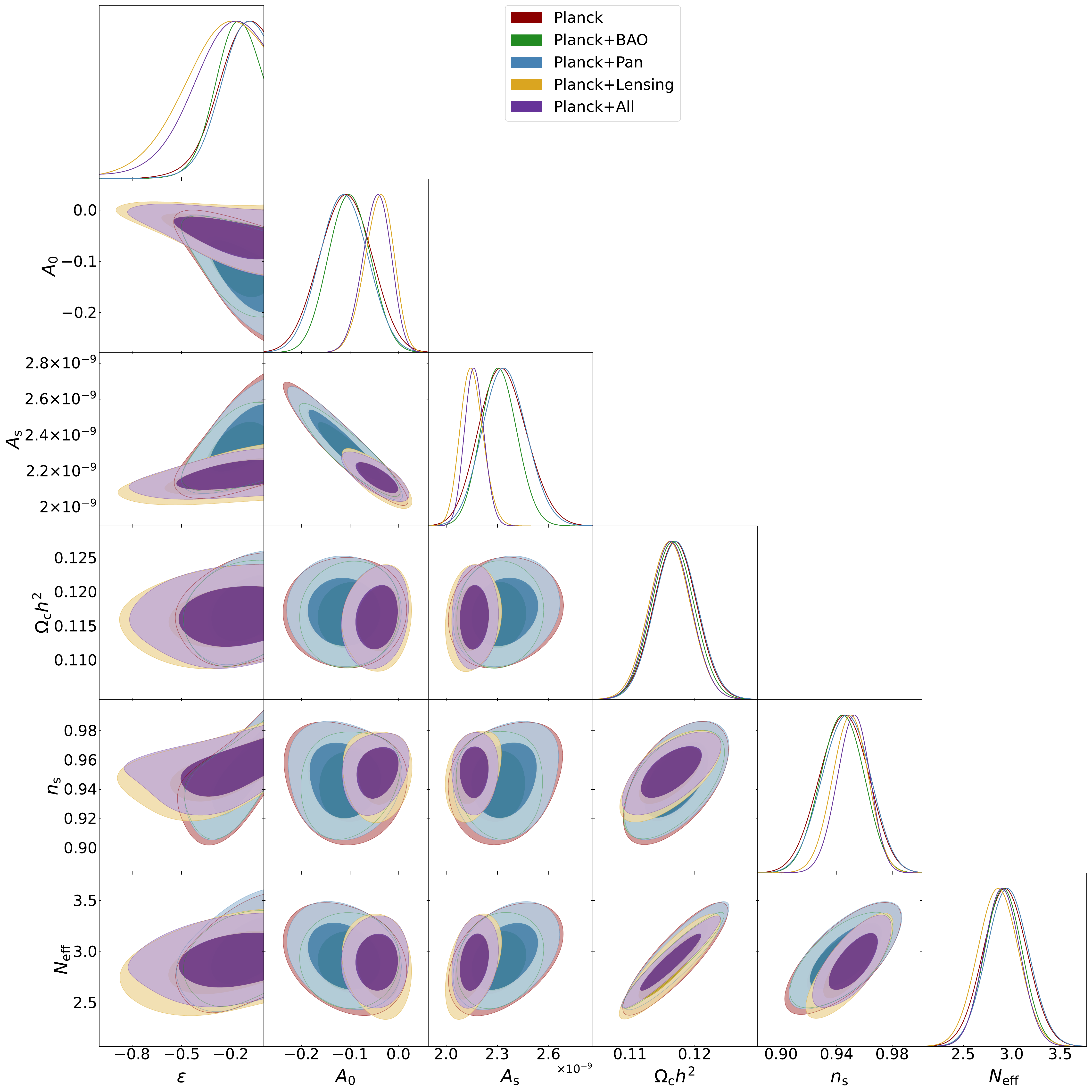}
	\caption{Marginalized 2D and 1D posterior distributions for the \slcdm$+ N_{\rm eff}$.}
	\label{fig:nnu}
\end{figure*}

%---------Neff+meff-----------%
\begin{table*}
	\begin{center}
		\renewcommand{\arraystretch}{1.5}
		\resizebox{0.9\textwidth}{!}{\begin{tabular}{c c c c  c  c}
  	        \hline
			\textbf{Parameter} & \textbf{Planck}  & \textbf{Planck+Lensing}  & \textbf{Planck+BAO} & \textbf{Planck+Pantheon} & \textbf{Planck+All} \\
			\hline\hline
  $\Omega_\mathrm{b} h^2$ &$ 0.02271\pm 0.00020$&$ 0.02260\pm 0.00018$&$ 0.02273\pm 0.00020$&$0.02272\pm 0.00019$&$0.02263\pm 0.00016$\\
$\Omega_\mathrm{c} h^2$ & $0.1186^{+0.0037}_{-0.0031}$&$0.1190^{+0.0037}_{-0.0030}$&$0.1180^{+0.0036}_{-0.0031}$&$0.1182^{+0.0038}_{-0.0032}$&$0.1181^{+0.0030}_{-0.0025}$\\
$100\theta_\mathrm{MC}$ & $1.04089\pm 0.00037$&$1.04085\pm 0.00036$&$1.04098\pm 0.00034$&$1.04095\pm 0.00036$&$ 1.04099\pm 0.00032$\\
$\tau_\mathrm{reio}$ &$ 0.0509\pm 0.0086$&$0.0511\pm 0.0087$&$ 0.0514\pm 0.0085$&$0.0515\pm 0.0083$&$0.0520\pm 0.0084$\\
$\log(10^{10} A_\mathrm{s})$ &$ 3.191\pm 0.056$&$3.097\pm 0.033$&$3.187\pm 0.051$&$3.193\pm 0.053$&$3.104\pm 0.031$\\
$n_\mathrm{s}$ &$ 0.954\pm 0.013$&$0.9598\pm 0.0093$&$0.955\pm 0.013$&$0.955\pm 0.013$&$0.9608\pm 0.0089$\\
\hline
$H_0$ [km s$^{-1}$ Mpc$^{-1}$] & $67.97\pm 0.97$&$67.81\pm 0.91$&$68.33^{+0.63}_{-0.78}$&$68.23^{+0.85}_{-0.97}$&$68.23^{+0.57}_{-0.63}$\\
$\sigma_8$ &$ 0.814\pm 0.036$&$0.789^{+0.028}_{-0.024}$&$0.824\pm 0.024$&$ 0.821^{+0.036}_{-0.032}$&$ 0.801^{+0.019}_{-0.017}$\\
$S_8$ &$ 0.834\pm 0.029$& $0.809^{+0.023}_{-0.020}$&$ 0.836\pm 0.023$&$ 0.836^{+0.031}_{-0.027}$&$ 0.812^{+0.019}_{-0.017}$\\
$A_{\rm 0}$ & $-0.140\pm 0.050$&$ -0.056\pm 0.034$&$-0.138\pm 0.046$&$-0.141\pm 0.047$&$-0.063\pm 0.033$\\
$\epsilon$ &$> -0.290$&$> -0.544$&$ > -0.279$&$ > -0.281$&$ > -0.481$\\
$N_{\rm eff}$&$ 3.21^{+0.10}_{-0.14}$&$3.183^{+0.089}_{-0.12}$&$3.176^{+0.088}_{-0.13}$&$3.190^{+0.096}_{-0.14}$&$3.143^{+0.066}_{-0.10}$\\
$m_{\rm eff}$ [eV] &$< 0.893$&$< 0.744$&$< 0.700$&$ < 0.801$&$< 0.571$\\          
\hline\hline
  \end{tabular}}
	\end{center}
	\caption{Results for \slcdm+$N_{\rm eff}+m_{\rm eff}$. The constraints on parameters are at $68\%$ CL, while upper bounds are at $95\%$ CL.}
	\label{tab:nnumeff}
\end{table*}

\begin{figure*}[htp]
	\centering
	\includegraphics[width=0.9 \textwidth]{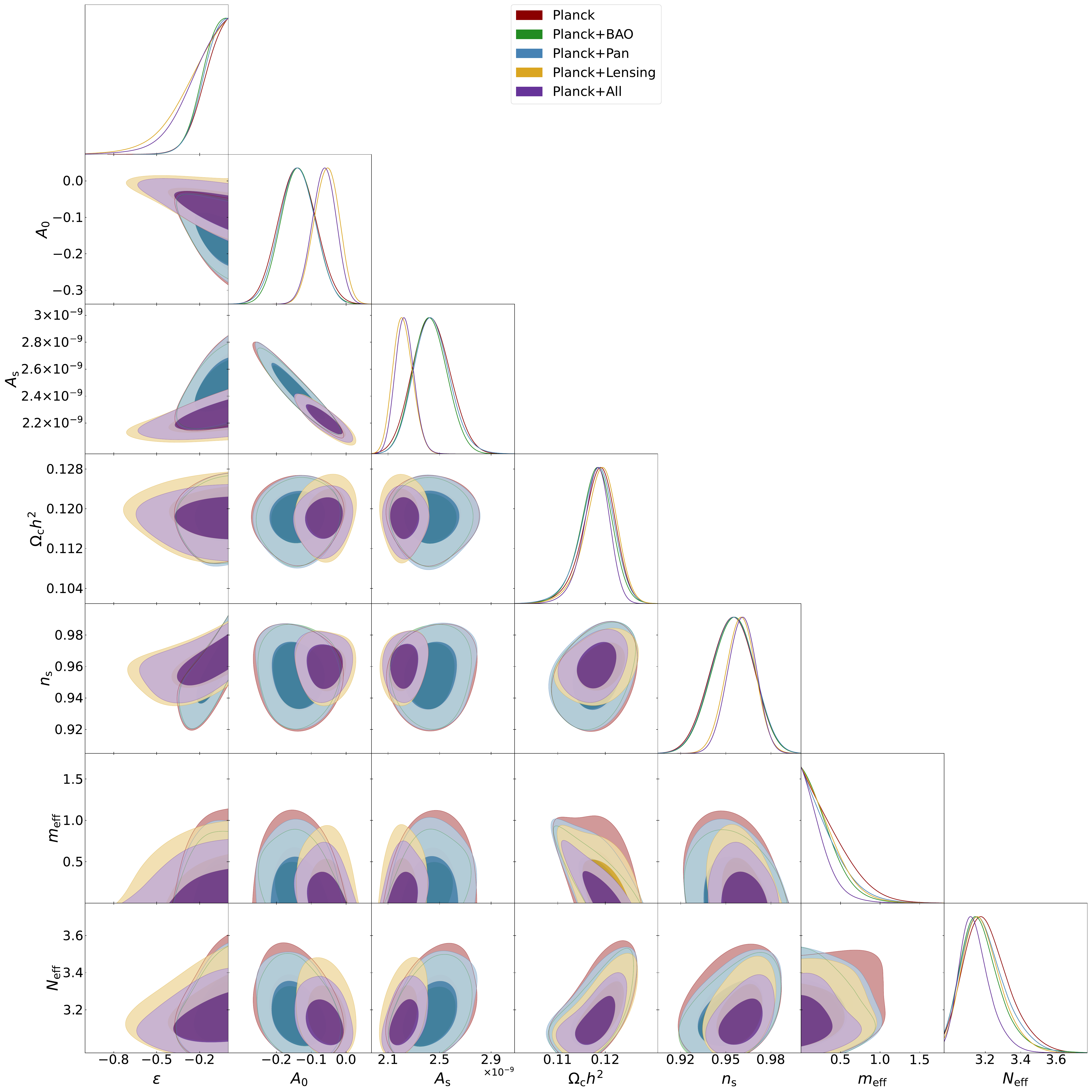}
	\caption{Marginalized 2D and 1D posterior distributions for the $\Lambda\rm CDM + N_{\rm eff}+m_{\rm eff}$.}
	\label{fig:nnumeff}
\end{figure*}

%---------mnu-----------%
\begin{table*}
	\begin{center}
		\renewcommand{\arraystretch}{1.5}
		\resizebox{0.9\textwidth}{!}{\begin{tabular}{c c c c  c  c}
  	        \hline
			\textbf{Parameter} & \textbf{Planck}  & \textbf{Planck+Lensing}  & \textbf{Planck+BAO} & \textbf{Planck+Pantheon} & \textbf{Planck+All} \\
			\hline\hline
$\Omega_\mathrm{b} h^2 $&$0.02254\pm 0.00018$&$ 0.02245\pm 0.00018$&$ 0.02259\pm 0.00015$&$0.02259\pm 0.00017$&$0.02254\pm 0.00015$\\
$\Omega_\mathrm{c} h^2$&$ 0.1185\pm 0.0015$&$0.1189\pm 0.0015$&$0.1182\pm 0.0013$&$0.1181\pm 0.0015$&$0.1182\pm 0.0012$\\
$100\theta_\mathrm{MC}$&$1.04104\pm 0.00033$&$1.04098\pm 0.00033$&$1.04111\pm 0.00030$&$1.04110\pm 0.00032$&$ 1.04110\pm 0.00030$\\
$\tau$& $0.0509\pm 0.0085$&$0.0507\pm 0.0083$&$0.0518\pm 0.0084$&$ 0.0520\pm 0.0085$&$ 0.0510\pm 0.0086$\\
$\ln(10^{10} A_\mathrm{s})$&$3.183\pm 0.055$&$3.102\pm 0.035$&$3.165\pm 0.052$&$3.177\pm 0.054$&$ 3.096^{+0.034}_{-0.038}$\\
$n_\mathrm{s} $&$0.949\pm 0.012$&$ 0.9555\pm 0.0088$&$0.952\pm 0.012$&$0.952\pm 0.012$&$0.9585\pm 0.0081$\\
\hline
$H_0 $ [km s$^{-1}$ Mpc$^{-1}$] &$66.5^{+2.0}_{-1.7}$&$ 66.4^{+1.7}_{-1.6}$&$67.71\pm 0.61$&$67.6^{+1.2}_{-1.0}$&$67.80\pm 0.57$\\
$\sigma_8$&$0.817^{+0.043}_{-0.036}$&$ 0.789^{+0.033}_{-0.028}$&$ 0.835\pm 0.019$&$0.836\pm 0.027$&$0.812^{+0.015}_{-0.013}$\\
$S_8$& $0.849\pm 0.022$&$ 0.823\pm 0.014$&$ 0.849\pm 0.018$&$ 0.852\pm 0.020$&$ 0.824\pm 0.012$\\
$A_{\rm 0}$&$-0.137\pm 0.051$&$-0.066\pm 0.037$&$-0.120\pm 0.049$&$-0.130\pm 0.051$&$-0.060^{+0.042}_{-0.036}$\\
$\epsilon$&$ > -0.302$&$ > -0.514$&$ > -0.340$&$ > -0.334$&$ > -0.545$\\
$\Sigma m_\nu$ [eV]&$ < 0.624$&$ < 0.542$&$ < 0.266$&$ < 0.356$&$ < 0.236$\\
\hline\hline
            
		\end{tabular}}
	\end{center}
	\caption{Results for \slcdm+$\Sigma m_\nu$. The constraints on parameters are at $68\%$ CL, while upper bounds are at $95\%$ CL.}
	\label{tab:mnu}
\end{table*}

\begin{figure*}[htp]
	\centering
	\includegraphics[width=0.9 \textwidth]{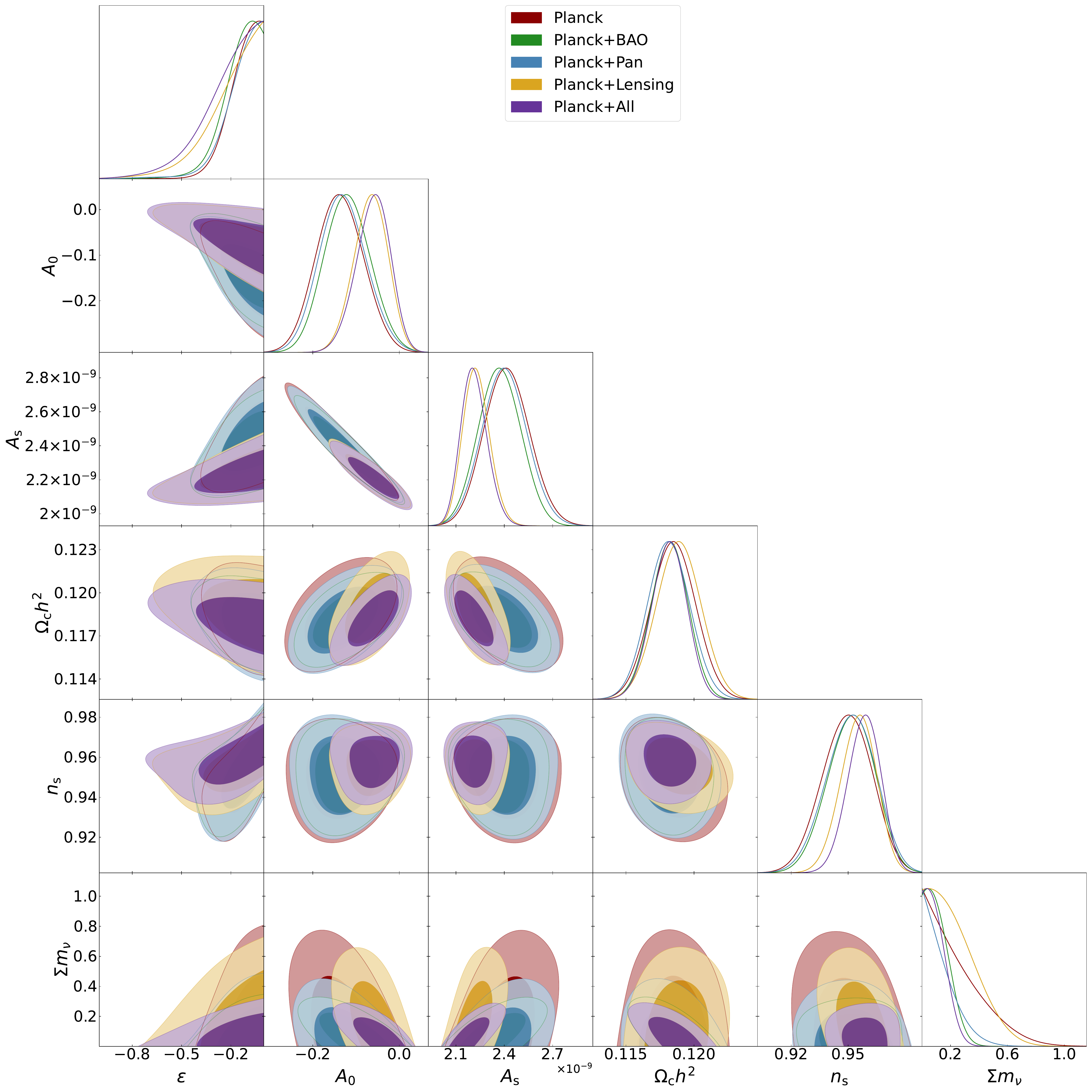}
	\caption{Marginalized 2D and 1D posterior distributions for the \slcdm$+\Sigma m_\nu$.}
	\label{fig:mnu}
\end{figure*}

%---------Neff+mnu-----------%
\begin{table*}
	\begin{center}
		\renewcommand{\arraystretch}{1.5}
		\resizebox{0.9\textwidth}{!}{\begin{tabular}{c c c c  c  c}
  	        \hline
			\textbf{Parameter} & \textbf{Planck}  & \textbf{Planck+Lensing}  & \textbf{Planck+BAO} & \textbf{Planck+Pantheon} & \textbf{Planck+All} \\
			\hline\hline
$\Omega_\mathrm{b} h^2$ &$0.02245\pm 0.00027$&$0.02230\pm 0.00025$&$0.02252\pm 0.00021$&$0.02255\pm 0.00024$&$ 0.02245\pm 0.00020$\\
$\Omega_\mathrm{c} h^2 $&$ 0.1171\pm 0.0031$&$0.1167\pm 0.0031$&$0.1170\pm 0.0031$&$0.1173\pm 0.0032$&$0.1166\pm 0.0030$\\
$100\theta_\mathrm{MC}$&$1.04121\pm 0.00047$&$1.04125\pm 0.00047$&$1.04128\pm 0.00046$&$1.04122\pm 0.00047$&$1.04131\pm 0.00046$\\
$\tau$&$ 0.0504\pm 0.0088$&$0.0503\pm 0.0083$&$0.0517\pm 0.0086$&$0.0513\pm 0.0086$&$0.0510\pm 0.0086$\\
$\ln(10^{10} A_\mathrm{s})$&$3.173^{+0.061}_{-0.070}$&$3.087^{+0.034}_{-0.039}$&$3.157\pm 0.049$&$3.175\pm 0.060$&$ 3.088^{+0.032}_{-0.038}$\\
$n_\mathrm{s}$&$0.943\pm 0.017$&$0.947\pm 0.013$&$0.946\pm 0.015$&$0.948\pm 0.016$&$0.953\pm 0.011$\\
\hline
$H_0$ [km s$^{-1}$ Mpc$^{-1}$] &$ 65.8\pm 2.2$&$65.3\pm 2.0$&$ 67.3\pm 1.2$&$67.2\pm 1.7$&$67.2\pm 1.1$\\
$\sigma_8$&$0.811^{+0.043}_{-0.038}$&$0.783^{+0.031}_{-0.027}$&$ 0.833\pm 0.019$&$0.834\pm 0.027$&$0.809\pm 0.014$\\
$S_8$ &$0.848\pm 0.022$&$ 0.822\pm 0.014$&$ 0.848\pm 0.018$&$ 0.852\pm 0.021$&$ 0.822\pm 0.012$\\
$A_{\rm 0}$&$-0.133\pm 0.056$&$ -0.058^{+0.038}_{-0.033}$&$-0.116\pm 0.044$&$-0.132\pm 0.052$&$-0.057^{+0.039}_{-0.032}$\\
$\epsilon$&$ > -0.367$&$> -0.591$&$ > -0.355$&$> -0.342$&$ > -0.587$\\
$N_{\rm eff} $&$2.94\pm0.22$&$ 2.88\pm 0.20$&$ 2.96\pm0.19$&$2.99\pm 0.21$&$2.94\pm0.19$\\
$\Sigma m_\nu$ [eV]&$ < 0.637$&$< 0.526$&$ < 0.254$&$< 0.365$&$< 0.225$\\
\hline\hline
            
		\end{tabular}}
	\end{center}
	\caption{Results for \slcdm+$N_{\rm eff}+\Sigma m_\nu$. The constraints on parameters are at $68\%$ CL, while upper bounds are at $95\%$ CL.}
	\label{tab:neffmnu}
\end{table*}

\begin{figure*}[htp]
	\centering
	\includegraphics[width=0.9 \textwidth]{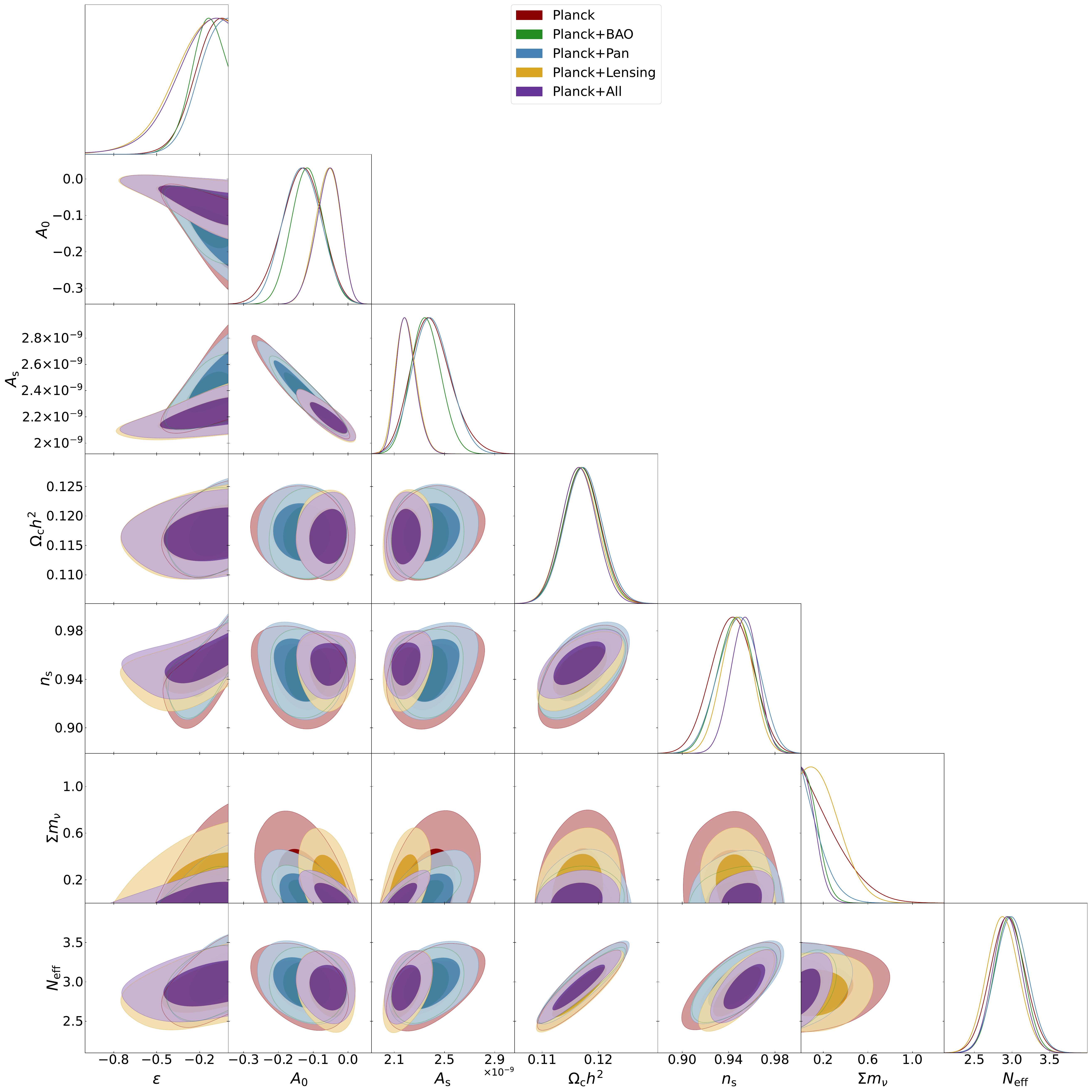}
	\caption{Marginalized 2D and 1D posterior distributions for the \slcdm$+ N_{\rm eff}+\Sigma m_\nu$.}
	\label{fig:nnumnu}
\end{figure*}

%---------Neff+meff+mnu-----------%
\begin{table*}
	\begin{center}
		\renewcommand{\arraystretch}{1.5}
		\resizebox{0.9\textwidth}{!}{\begin{tabular}{c c c c  c  c}
  	        \hline
			\textbf{Parameter} & \textbf{Planck}  & \textbf{Planck+Lensing}  & \textbf{Planck+BAO} & \textbf{Planck+Pantheon} & \textbf{Planck+All} \\
			\hline\hline
$\Omega_\mathrm{b} h^2$&$0.02268\pm 0.00021$&$0.02255\pm 0.00019$&$0.02273\pm 0.00018$&$0.02273\pm 0.00019$&$0.02263\pm 0.00017$\\
$\Omega_\mathrm{c} h^2$&$ 0.1186^{+0.0038}_{-0.0032}$&$0.1191^{+0.0035}_{-0.0029}$&$0.1177^{+0.0035}_{-0.0030}$&$0.1180^{+0.0038}_{-0.0032}$&$0.1182^{+0.0030}_{-0.0025}$\\
$100\theta_\mathrm{MC} $&$ 1.04083\pm 0.00038$&$1.04081\pm 0.00037$&$1.04100\pm 0.00034$&$1.04094\pm 0.00036$&$1.04098\pm 0.00033$\\
$\tau_\mathrm{reio} $&$0.0507\pm 0.0084$&$0.0507\pm 0.0087$&$0.0514\pm 0.0088$&$0.0515\pm 0.0086$&$0.0515\pm 0.0085$\\
$\log(10^{10} A_\mathrm{s})$&$3.210\pm 0.059$&$3.117^{+0.039}_{-0.044}$&$3.199\pm 0.057$&$3.204\pm 0.056$&$3.109\pm 0.034$\\
$n_\mathrm{s}$&$ 0.952\pm 0.014$&$0.9583\pm 0.0099$&$ 0.955\pm 0.013$&$0.955\pm 0.014$&$0.9612\pm 0.0093$\\
\hline
$H_0$ [km s$^{-1}$ Mpc$^{-1}$] &$66.6^{+1.9}_{-1.7}$&$66.7\pm 1.5$&$68.18\pm 0.73$&$67.8\pm 1.2$&$68.18\pm 0.67$\\
$\sigma_8$&$ 0.787\pm 0.044$&$0.771\pm 0.030$&$0.822\pm 0.025$&$0.813\pm 0.035$&$0.800^{+0.020}_{-0.018}$\\
$S_8$ &$0.828\pm 0.029$&$ 0.809^{+0.022}_{-0.019}$&$ 0.835\pm 0.023$&$ 0.834\pm 0.028$&$ 0.812^{+0.019}_{-0.017}$\\
$A_{\rm 0} $&$-0.156\pm 0.052$&$-0.076\pm 0.041$&$-0.148\pm 0.051$&$ -0.151\pm 0.050$&$ -0.070\pm 0.034$\\
$\epsilon$&$ > -0.264$&$ > -0.457$&$ > -0.277$&$ > -0.268$&$> -0.431$\\
$N_{\rm eff}$ &$3.21^{+0.11}_{-0.15}$&$3.177^{+0.088}_{-0.13}$&$3.168^{+0.083}_{-0.12}$&$3.188^{+0.095}_{-0.14}$&$3.150^{+0.072}_{-0.11}$\\
$\Sigma m_\nu $ [eV]&$< 0.593$&$ < 0.483$&$ < 0.234$&$< 0.335$&$< 0.198$\\
$m_{\rm eff}$ [eV]&$ < 0.909$&$< 0.695$&$< 0.689$&$< 0.790$&$ < 0.550$\\
\hline\hline
		\end{tabular}}
	\end{center}
	\caption{Results for \slcdm+$N_{\rm eff}+m_{\rm eff}+\Sigma m_\nu$. The constraints on parameters are at $68\%$ CL, while upper bounds are at $95\%$ CL.}
	\label{tab:nnumeffmnu}
\end{table*}

\begin{figure*}[htp]
	\centering
	\includegraphics[width=0.9 \textwidth]{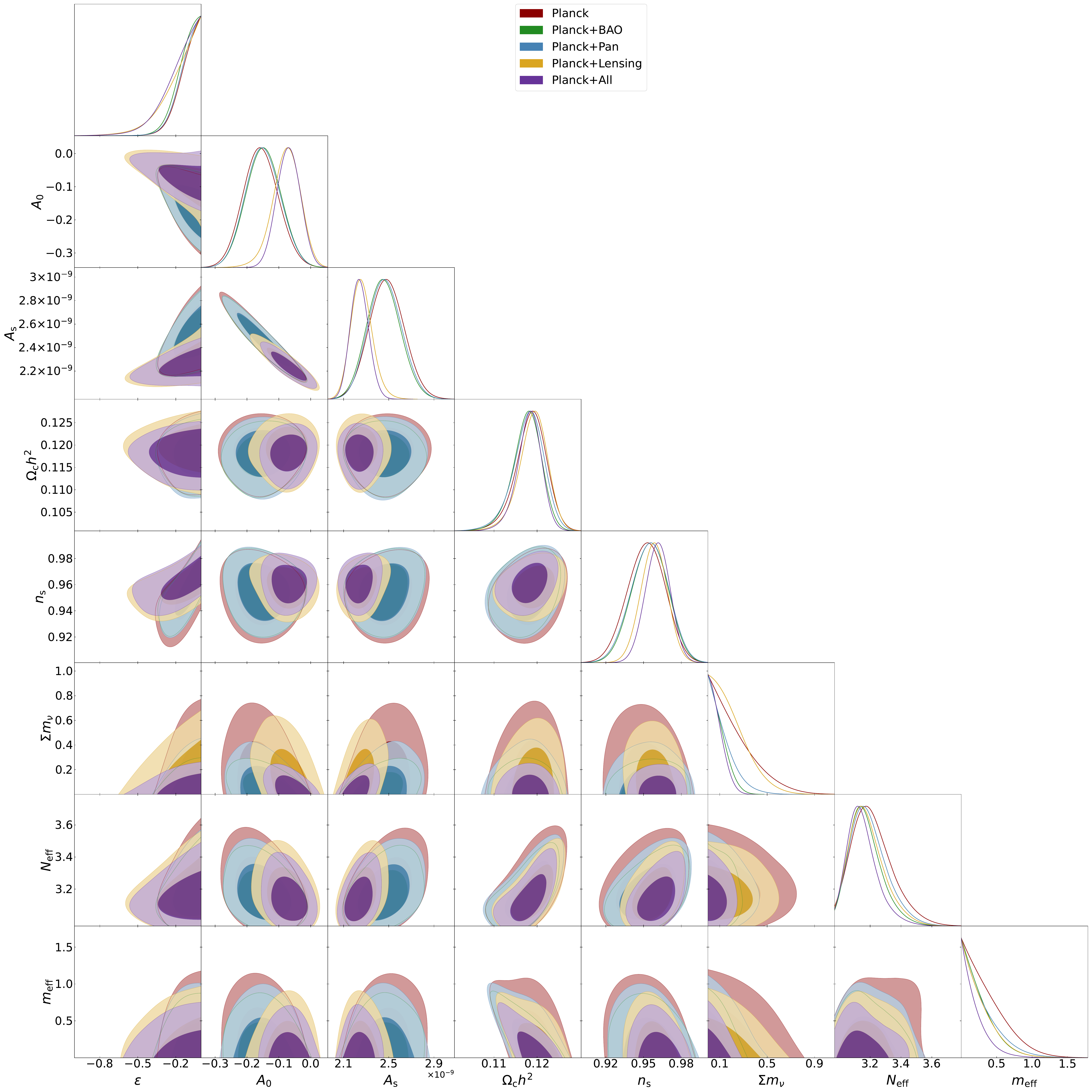}
	\caption{Marginalized 2D and 1D posterior distributions for the \slcdm$+ N_{\rm eff}+m_{\rm eff}+\Sigma m_\nu$.}
	\label{fig:nnumeffmnu}
\end{figure*}
\begin{figure}[htp]
	\centering
	\includegraphics*[width=15pc]{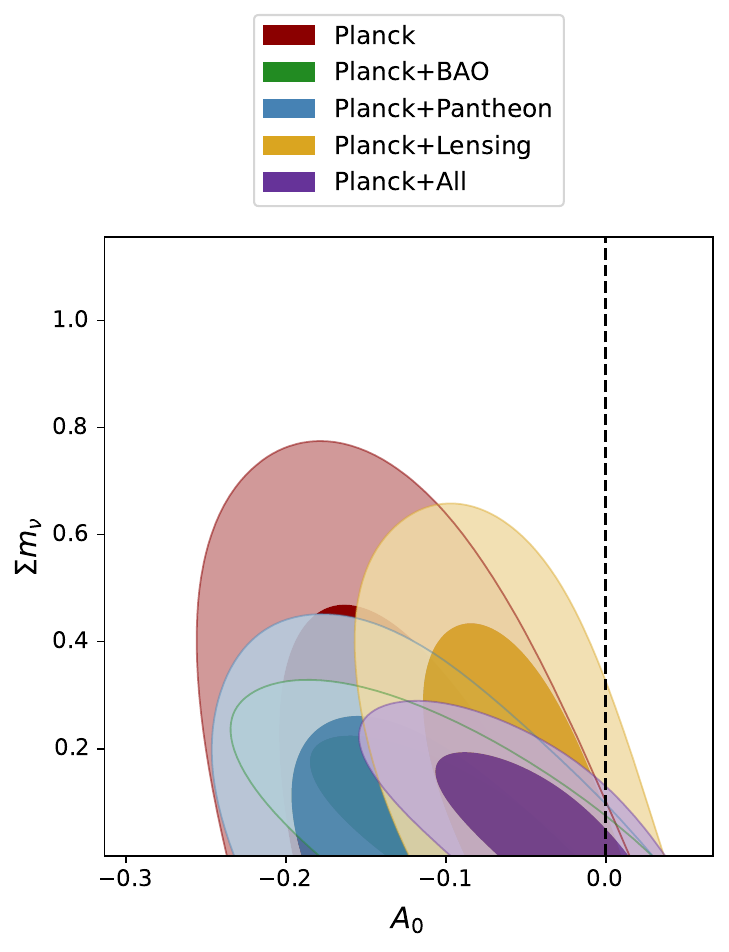}
	\caption{Marginalized 2D posterior distributions for the \slcdm+$\mnu$.}
	\label{fig:A0mnu}
\end{figure}

	\label{fig:meffNeffmnu_A0As}

\section{Results} \label{results}

Here, we list the constraints on Super-$\Lambda$CDM and its various extensions that we obtained with the MCMC analyses. Full results can be found in~\tab{lcdm} and \fig{lcdm} (for only Super-$\Lambda$CDM),~\tab{nnu} and \fig{nnu} (Super-$\Lambda$CDM$+\nnu$)~\tab{nnumeff} and \fig{nnumeff} (Super-$\Lambda$CDM$+\nnu+\meff$),~\tab{mnu} and \fig{mnu} (Super-$\Lambda$CDM$+\mnu$),~\tab{neffmnu} and \fig{nnumnu} (Super-$\Lambda$CDM$+\nnu+\mnu$), and ~\tab{nnumeffmnu} with \fig{nnumeffmnu} (Super-$\Lambda$CDM$+\nnu+\mnu+\meff$). First of all, it should be stressed that, as pointed out in~\cite{Adhikari:2019fvb,Adhikari:2022moo}, the BAO and Lensing datasets have to be taken \textit{cum grano salis}\footnote{From Latin, it translates into "with a pinch of salt" and means that care is needed.} as theoretical prediction for the Super-$\Lambda$CDM model is still under development. However, using an agnostic approach, we include both in our exploration of the parameter spaces. On the other hand, supernovae data in the Pantheon measurements are not affected by the primordial NG and can be safely included in our discussion. The label Planck+All  refers to the combination of all the datasets (``All'' means ``Lensing+BAO+Pantheon''). We present the results quoting lower bounds at $95\%$ CL and constraints at $68\%$ CL, if not otherwise stated. For the sake of simplicity we refer to $\meff$ as $\mef$.

\subsubsection{Super-parameters}

In presenting the constraints on $A_0$, we should bear in mind that $A_0=0$ corresponds to the standard $\Lambda$CDM scenario with no contribution from NG at large scale. Therefore, any deviation of $A_0$ from its null value will go in favor of Super-$\Lambda$CDM. 

Before exploring the neutrino sector, we note that the polarization did not change much the constraints obtained in~\cite{Adhikari:2019fvb}. In fact, with only TT+$\tau$ prior (with the Planck 2015 release), $A_0=-0.15^{+0.14}_{-0.13}$ at \ncl~\cite{Adhikari:2019fvb} which increases when combined with distance ladder $H_0$ and SNe Ia Pantheon leading to  $A_0=-0.21^{+0.12}_{-0.13}$ at \ncl~\cite{Adhikari:2019fvb}. In our case, for Planck, we obtain $A_0=-0.12\pm0.10$ and for Planck+Pantheon, $A_0=-0.118^{+0.095}_{-0.092}$, both at \ncl. Despite the absolute value being lower than the previous analysis~\cite{Adhikari:2019fvb}, it is better constrained. On the other hand, if we allow the possibility of different relativistic species in the universe, by allowing $\nnu$ to vary, the mean value shifts toward $0$. For example, with Planck only we have  $A_{\rm 0} = -0.109^{+0.095}_{-0.098}$ at \ncl. The highest absolute value is obtained when we promote as free parameters not only $\nnu$ but also $\mef$ and $\mnu$. In particular,  for Planck only, we get $A_{\rm 0} = -0.16 \pm 0.11$ at \ncl. 
Interestingly enough, combinations of Planck data with Lensing or BAO leads to the least solid evidence for NG. For example, Planck+Lensing predicts the most compatible value with zero for $A_0$, leading to 
$A_{\rm 0} = -0.040^{+0.055}_{-0.060}$
at \ncl with $\nnu$ as free parameter (i.e. for the scenario Super-$\Lambda$CDM$+\nnu$). The latter result hints that NG are negligible at less than $2\sigma$. Nonetheless, apart from the Lensing combination, our predictions for a negative nonzero $A_0$ lies within $\sim 2.5\sigma$. The lensing dataset affects greatly the Planck+All combination, and, therefore, $A_0$ is shifted toward zero, thus increasing the compatibility with $\Lambda$CDM with respect to other datasets.

In \slcdm, $A_0$ negatively correlates with the primordial amplitude $A_s$ and there is a hint of correlation with $\Omega_c h^2$, as can be seen in the triangular plot in \fig{lcdm}. When considering extensions to the neutrino sector, the degeneracy with $A_s$ persists (and is even more pronounced, e.g., see \fig{nnumeffmnu}), while the correlation with $\Omega_c h^2$ weakens or it is completely absent (see, for example, \fig{nnu}).

Concerning the second Super-parameter $\epsilon$,
it has to be underlined that the usual trispectrum for multifield inflationary model is recovered for $\epsilon=0$. That is, we impose a cutoff at $0$ as indicated in \tab{Priors}. 
In our analysis, we did not find corroborating evidence for a nonzero value of $\epsilon$. Instead, we put lower bounds at \ncl which are generally more relaxed than the previous analysis, where it was found $\epsilon>-0.320$ (for TT+$\tau$ prior) and $\epsilon>-0.200$ when late-time measurements were included~\cite{Adhikari:2019fvb}. Highest absolute values are possible when the lensing data are included. In fact, for \slcdm+$\nnu$, we have relaxed bounds, such as $\epsilon>-0.692$ and $\epsilon>-0.388$ for Planck+Lensing and Planck+Pantheon, respectively. In all triangular plots, it is evident that Planck, Planck+Pantheon and Planck+BAO exhibit a certain correlation between $n_s$ and $\epsilon$, as we might expect due to the shift in the spectral index in \eq{CAMBRelation}. Conversely, the datasets, Planck+Lensing and Planck+All display a less pronounced degeneracy.

\subsubsection{Neutrino parameters}

Hints for additional massless particles are described by the parameter $N_{\rm eff}$ whose standard value is approximately $N_{\rm eff}\approx 3.044$. 
In our analysis, when we allow $N_{\rm eff}$ to vary, we see similar constraints as in the simple extension $\Lambda$CDM$+N_{\rm eff}$~\cite{Planck:2018vyg} for Planck only, which is $N_{\rm eff}=2.92\pm0.19$ at \scl, whereas in \slcdm+$\nnu$, $N_{\rm eff}=2.93\pm0.21$ at \scl. Therefore, Planck alone does not alter the constraints on $N_{\rm eff}$. 
The addition of BAO induces a preference for smaller values of $N_{\rm eff}$ in the \slcdm scenario leading to $N_{\rm eff}=3.01\pm0.18$ (at \scl for $\Lambda$CDM$+N_{\rm eff}$) and $N_{\rm eff}=2.91\pm0.18$ (at \scl for \slcdm+$\nnu$). 
Thus, the standard value remains within the $1\sigma$ range but deviation from it of more than $20\%$ is allowed at $95\%$~CL. In a similar way, when the sum of neutrino masses $\mnu$ is let free to vary, analogous consideration can be drawn where for Planck alone,  $N_{\rm eff}=2.91\pm0.19$ at \scl (for $\Lambda$CDM$+\nnu+\mnu$)~\cite{Planck:2018vyg}
and $N_{\rm eff}=2.94 \pm 0.22$ at 68\% CL for the \slcdm$+\nnu+\mnu$  model. 
Conversely, as soon as the possibility of a sterile neutrino is added to the model instead of $\mnu$, we obtain higher values for all datasets but 
the standard value still lies within the 95\% CL. Likewise, if we consider all three neutrino parameters in our model, i.e. for the model Super-$\Lambda$CDM$+\nnu+\mnu+\meff$, we have a preference for a higher effective number of relativistic species. In fact, when all the datasets are combined, we obtain $N_{\rm eff}=3.150^{+0.072}_{-0.11}$ at \scl (for Super-$\Lambda$CDM$+\nnu+\mnu+\meff$) while for Planck alone, $N_{\rm eff}=3.21^{+0.11}_{-0.15}$ at \scl (for Super-$\Lambda$CDM$+\nnu+\mnu+\meff$). Notably, these values are higher than the predictions of the $\Lambda$CDM model. For the same datasets (Planck+All and Planck), at \scl, we obtain $N_{\rm eff}=3.133^{+0.061}_{-0.099}$ and $N_{\rm eff}=3.156^{+0.074}_{-0.12}$ respectively.

The possibility of ruling out the inverted mass hierarchy is far from being reached in this analysis. Nonetheless, it is worth highlighting that the constraints are considerably weaker compared to those of the $\Lambda$CDM model. It was the case for $N_{\rm eff}$ but it is more evident with the sum of neutrino masses. The reason for the weaker constraints is not only due to a volume effect, but a genuinely negative correlation between $A_0$ and $\Sigma m_\nu$, as we can see in the triangular plots and in \fig{A0mnu}. In other words, given that there is a slight indication for $A_0<0$ coming out from our analysis, this translates to more room for massive neutrinos. Fixing $A_0=0$ could, therefore, bias the constraints on the total neutrino mass on being too strong. For example, in \slcdm with Planck only and Planck+BAO we have $\Sigma m_\nu<0.624\ev$ and $\Sigma m_\nu<0.266\ev$ at 95\% CL, respectively, while the $\lcdm$ predictions are $\Sigma m_\nu<0.257\ev$ and $\mnu<0.126\ev$ at 95\% CL. 
Because of the absence of noticeable correlation with the possibility of massless relics (i.e. $\Delta N_{\rm eff}\neq 0$), when $N_{\rm eff}$ is added to the set, the constraints remain almost unchanged. If also the sterile neutrino is included in the picture, the upper bounds are slightly stronger as, for example, for Planck only we have $\Sigma m_\nu<0.593\ev$ at 95\% CL. These results preserve the weaknesses of the constraining power of \slcdm compared to $\lcdm$ (as for Planck only we have, $\Sigma m_\nu<0.352\ev$ at 95\% CL). Additionally, it is noteworthy that the constraints on the neutrino mass sum get more relaxed when the analysis is limited to early Universe physics, in contrast to scenarios confined to low-redshift cosmology. For example, in comparison with our result for \slcdm+$\mnu$ using only Planck data, in~\cite{Giare:2023aix} they found an increase of $~21\%$ in the constraints, when lensing, BAO and type Ia Supernovae data are combined, without relying on the CMB.
%\textbf{\color{violet}The possibility of a bias in constraining the total neutrino mass persists even when we assume a similarity of approximately $1$, so there is a nearly perfect alignment with the local-shape and the constraints are now relevant. Although the upper bound for $\Sigma m_\nu$ is lower, it is still not as stringent as in the scenario where NG are not considered. For example if we let only $\Sigma m_\nu$ to vary the bound  $\Sigma m_\nu<0.617$~eV shift towards $\Sigma m_\nu<0.569$~eV as we apply the constraints on the trispectrum. These results can be seen in \fig{mnu_tauprior}.}

Similar results are obtained for $\mef$. In the scenario where massive sterile neutrinos are combined with the standard active neutrinos, we exclude the possibility of considering these neutrinos as candidates for cold dark matter particles. To be specific, we apply a cutoff at thermal masses greater than $10$~eV. Again, we obtain weaker constraints with respect to $\lcdm$~\cite{Planck:2018vyg}. For example, in $\lcdm+\nnu+\mef$, Planck alone gives $\mef<0.753\ev$ at 95\% CL and it increases to $\mef<0.893\ev$ at 95\% CL considering \slcdm. In \slcdm, the upper bound is stronger when Planck is combined with Pantheon, i.e. $\mef<0.801\ev$ at 95\% CL. If we add also the sum of neutrino masses, the bound is again stronger not only for Planck+Pantheon $\mef<0.790\ev$ at 95\% CL, but also for other datasets except for Planck alone, where the limit relaxes to $\mef<0.909\ev$ at 95\% CL. For comparison, Planck alone in $\Lambda$CDM$+\mef+\nnu+\mnu$ at \ncl gives $\mef<0.339\ev$.
This is an interesting observation in this context which dictates that the bounds on the neutrino mass obtained in \slcdm
are relaxed compared to the $\Lambda$CDM paradigm. 

\subsubsection{$H_0$ and $S_8$}

Concerning the cosmological tensions, we can see that, adding the Super-parameter $A_0$,  the value of $H_0$  increases a bit with respect to the $\Lambda$CDM cosmology. In $\Lambda$CDM we have $H_0=67.27\pm 0.60 \,\mathrm{km}\, \mathrm{s}^{-1}\,\mathrm{Mpc}^{-1}$ at \scl that becomes $H_0=68.13\pm0.71 \, \mathrm{km}\, \,\mathrm{s}^{-1}\, \mathrm{Mpc}^{-1}$ at \scl in \slcdm.  
It is, however, smaller than the one found in the previous work with CMB temperature data only~\cite{Adhikari:2019fvb}. A similar increase with the inclusion of the Super-parameters is seen also within the extension of the neutrino phenomenology. For instance with $\nnu$ we have $H_0=66.4\pm1.4 \,\mathrm{km}\, \mathrm{s}^{-1}\,\mathrm{Mpc}^{-1}$ (at \scl) for $\lcdm+\nnu$  that changes to  $H_0=67.3 \pm 1.7 \mathrm{km}\, \mathrm{s}^{-1}\,\mathrm{Mpc}^{-1}$ (at \scl) in \slcdm +$\nnu$. 
Moving into \slcdm tends to ease the tension with late-time measurements, which, however, are still well beyond $3\sigma$. 

If we let the effective mass of sterile neutrinos free to vary, we see that $A_0$ increases and $H_0$ increases accordingly,  while the $S_8$ values decrease. For example, with Planck only data the \scl constraint on $H_0$ is $H_0=67.97 \pm 0.97\; \mathrm{km}\, \mathrm{s}^{-1}\,\mathrm{Mpc}^{-1}$. On the other hand, if we consider the sum of the neutrino masses, due to the decrease in its expectation values as well as its inverse proportionality with $H_0$, we obtain a lower value for $H_0$. For instance, we have $H_0=67.0^{+1.2}_{-0.97}\, \mathrm{km}\, \mathrm{s}^{-1}\,\mathrm{Mpc}^{-1}$ and $H_0=66.5^{+2.0}_{-1.7}\, \mathrm{km}\, \mathrm{s}^{-1}\,\mathrm{Mpc}^{-1} $  both at \scl for Planck only $\Lambda$CDM+$\mnu$ and \slcdm+$\mnu$ respectively.
If we promote the effective number of relativistic species as a parameter of the theory, i.e. \slcdm+$\mnu$+$\nnu$, with $A_0=-0.133\pm0.056$, $H_0$ decreases as much as (with Planck only) $H_0=65.8\pm2.2 \, \mathrm{km}\, \mathrm{s}^{-1}\,\mathrm{Mpc}^{-1}$ at \scl.
The latter value is greater than the one corresponding to $A_0=0$ (which at \ncl is $66.1^{+3.5}_{-3.6} \, \mathrm{km}\, \mathrm{s}^{-1}\,\mathrm{Mpc}^{-1}$.
When all three neutrino parameters are included in the analysis (i.e., for the scenario Super-$\Lambda$CDM$+\nnu+\mnu+\meff$), we get $H_0=68.18\pm 0.73 \, \mathrm{km}\, \mathrm{s}^{-1} \mathrm{Mpc}^{-1}$ at 68\% CL for Planck+BAO and $H_0 = 67.8\pm1.2 \,\mathrm{km}\, \mathrm{s}^{-1} \mathrm{Mpc}^{-1}$ at 68\% CL for Planck+Pantheon. The corresponding values for the same model with null $A_0$, at \scl, are $H_0=68.0^{+0.67}_{-0.81}\, \mathrm{km}\, \mathrm{s}^{-1} \mathrm{Mpc}^{-1}$ and $H_0 = 67.53\pm0.97 \,\mathrm{km}\, \mathrm{s}^{-1} \mathrm{Mpc}^{-1}$.
Thus, looking at the estimated values of $H_0$ in \slcdm and its various extensions where the maximum mean value of $H_0$ is $\sim 68.33\; \mathrm{km}\, \mathrm{s}^{-1} \mathrm{Mpc}^{-1}$  (obtained in Super-$\Lambda$CDM$+\nnu+\meff$) with uncertainties less than $1 \,\mathrm{km}\, \mathrm{s}^{-1} \mathrm{Mpc}^{-1}$, one can conclude that neither \slcdm, nor its extensions, are efficient in resolving the $H_0$ tension between Planck~\cite{Planck:2018vyg} and SH0ES~\cite{Riess:2021jrx}. On the other hand, focusing on the 
estimated values of $S_8$ in \slcdm and its various extensions, we see that  compared to the $\Lambda$CDM-based Planck's estimation ($S_8 =  0.834 \pm 0.016$)~\cite{Planck:2018vyg}, the values of $S_8$ do not make any dramatic changes except in the case for Planck+Lensing for which a mild reduction in the $S_8$ parameter is observed for Super-$\Lambda$CDM$+\nnu+\meff$ ($S_8 = 0.809^{+0.023}_{-0.020}$ at 68\% CL),  Super-$\Lambda$CDM$+\nnu+\mnu+\meff$ ($S_8 = 0.809^{+0.022}_{-0.019}$ at 68\% CL). However, for other datasets, $S_8$ takes $\Lambda$CDM-like values or even higher. 
Therefore, it is clear that easing tension on $S_8$ in these scenarios  does not seem promising at least according to the current level of sensitivity of the astronomical data, even if for a fair comparison we should analyze the weak lensing data only assuming the Super-$\Lambda$CDM model as well.  
As no dataset constrains $\epsilon$ in such a way to have stronger bonds for $\tau$ we have not reported the bounds on $\tau$ for every dataset in the table and in the plots.
\begin{figure*}[htb]
    \centering
    \begin{subfigure}[b]{0.48\textwidth}
        \centering
        \includegraphics[width=\textwidth]{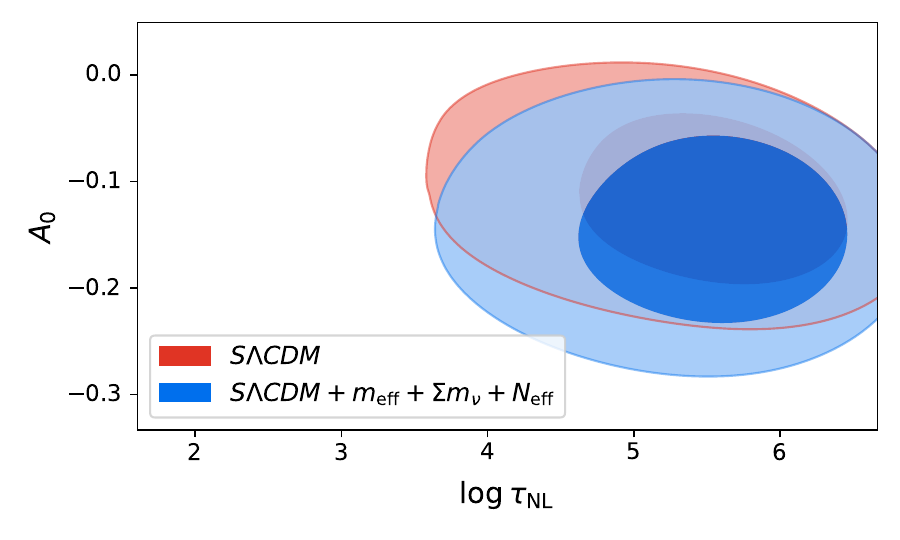}
        \label{fig:a0-vs-tau}
    \end{subfigure}
    \hfill
    \begin{subfigure}[b]{0.48\textwidth}
        \centering
        \includegraphics[width=\textwidth]{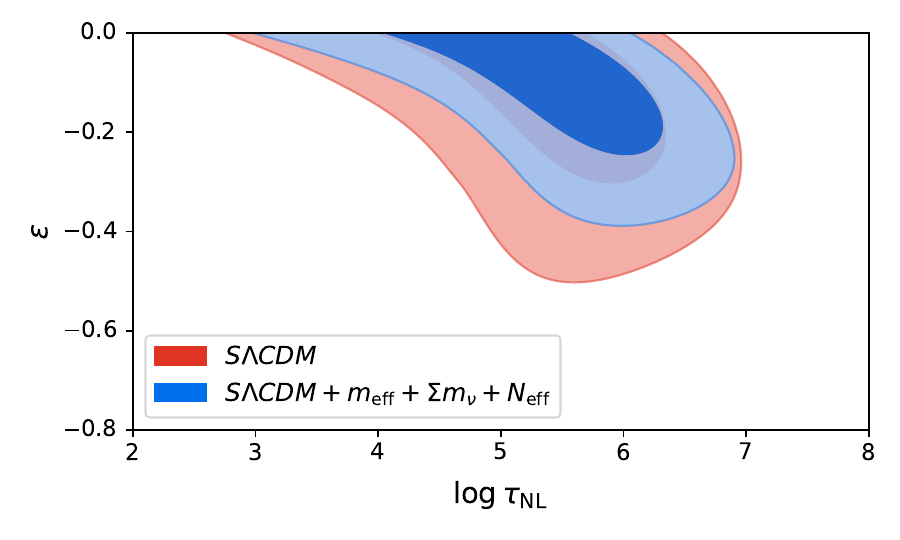}
        \label{fig:eps-vs-tau}
    \end{subfigure}
    \caption{$\tau_{\rm NL}$ is weakly constrained as we set only a lower bound for $\epsilon$. Here, only the Planck dataset has been used.}
    \label{fig:taunl}
\end{figure*}

\begin{figure}[htp]
	\centering
	\includegraphics[width=0.5\textwidth]{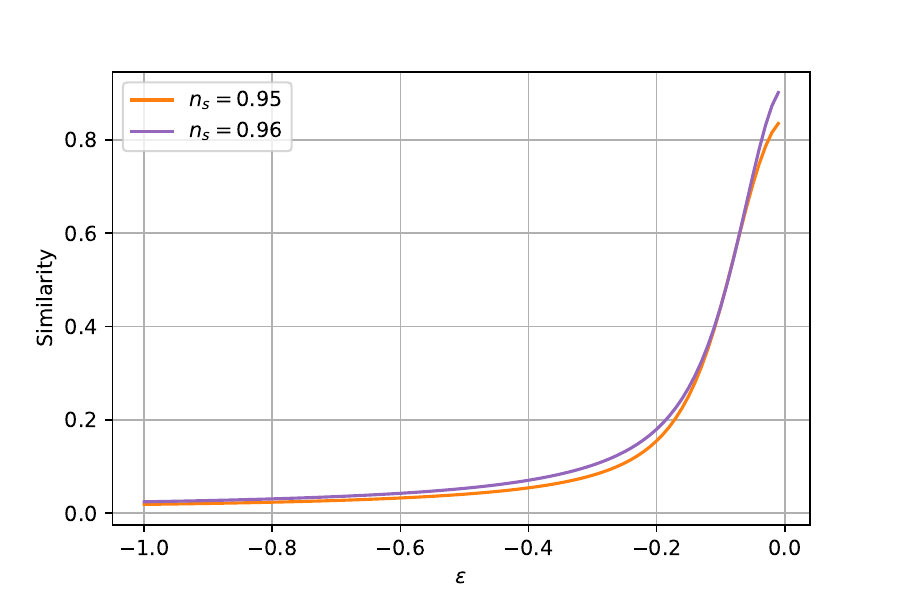}
	\caption{
 Cosine similarity between \eq{trispectrum} and the standard template. When $\epsilon\sim0$ we have a near overlap: The \textit{intermediate} shape tends to the local one, where the bounds are applicable.}
	\label{fig:similarity}
\end{figure}

\begin{figure*}[htp]
    \centering
    \begin{subfigure}[b]{0.32\textwidth}
        \centering
        \includegraphics[width=\textwidth]{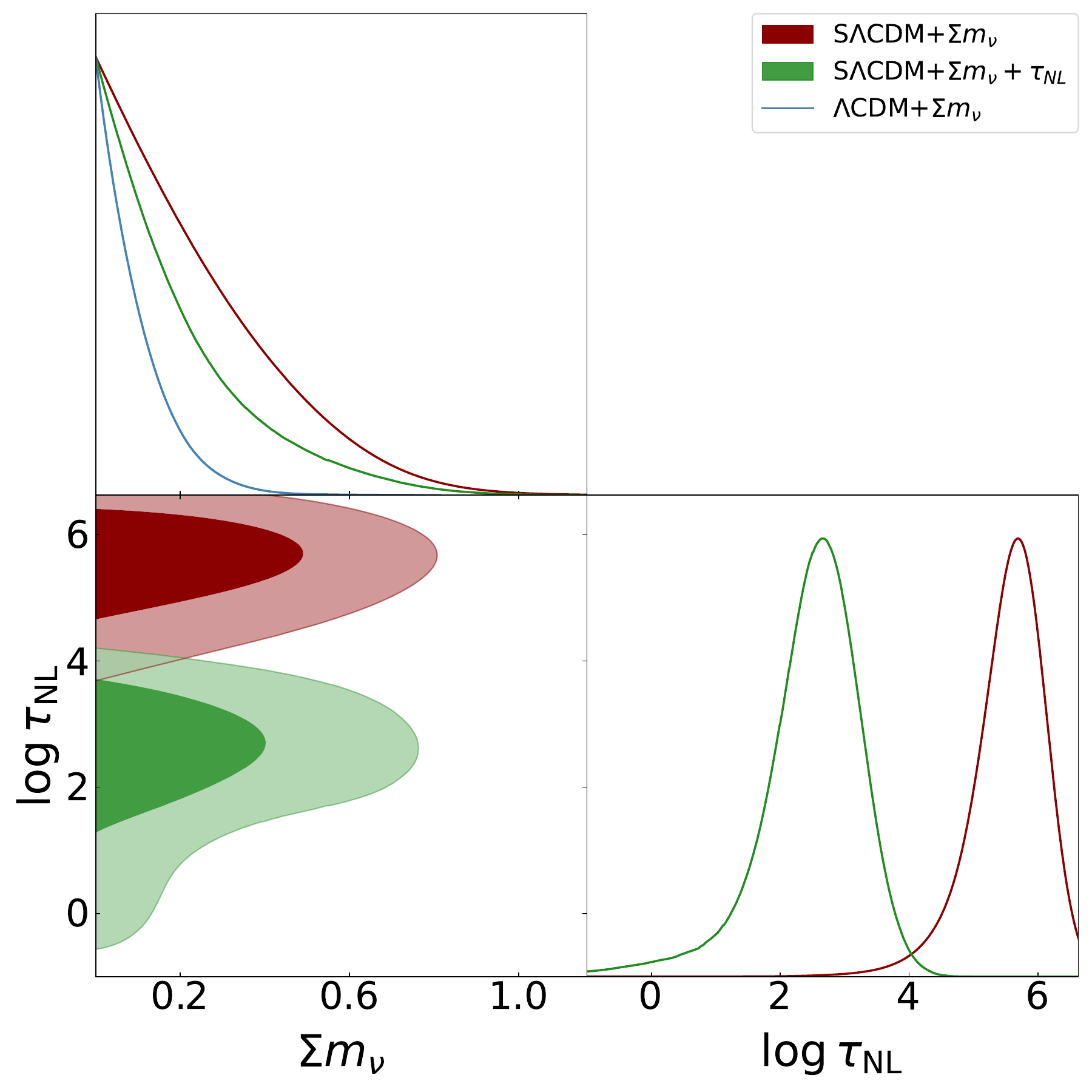}
        \caption{$\Sigma m_\nu$}
        \label{fig:1}
    \end{subfigure}
    \hfill
    \begin{subfigure}[b]{0.32\textwidth}
        \centering
        \includegraphics[width=\textwidth]{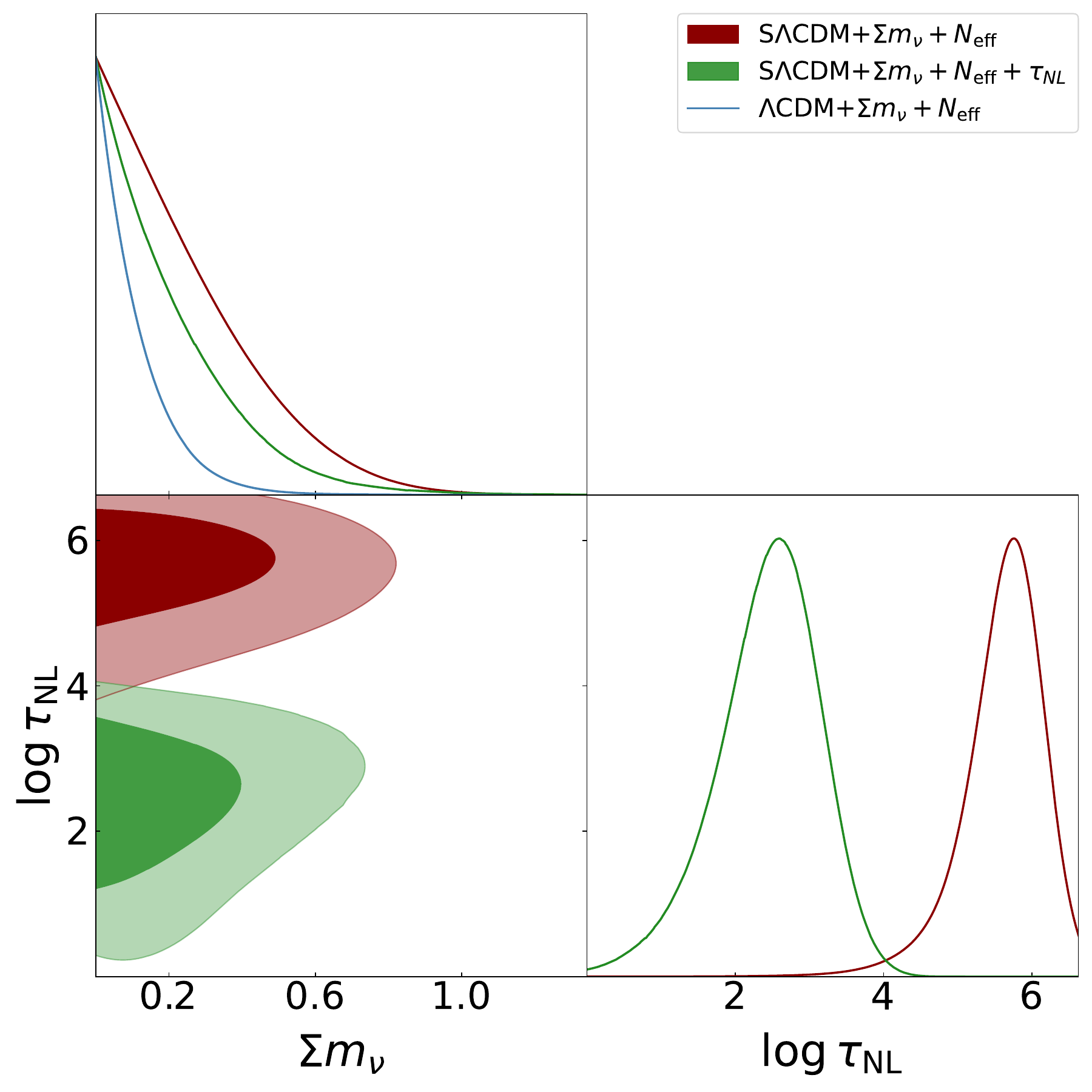}
        \caption{$\Sigma m_\nu+N_{\rm eff}$}
        \label{fig:2}
    \end{subfigure}
    \hfill
    \begin{subfigure}[b]{0.32\textwidth}
        \centering
        \includegraphics[width=\textwidth]{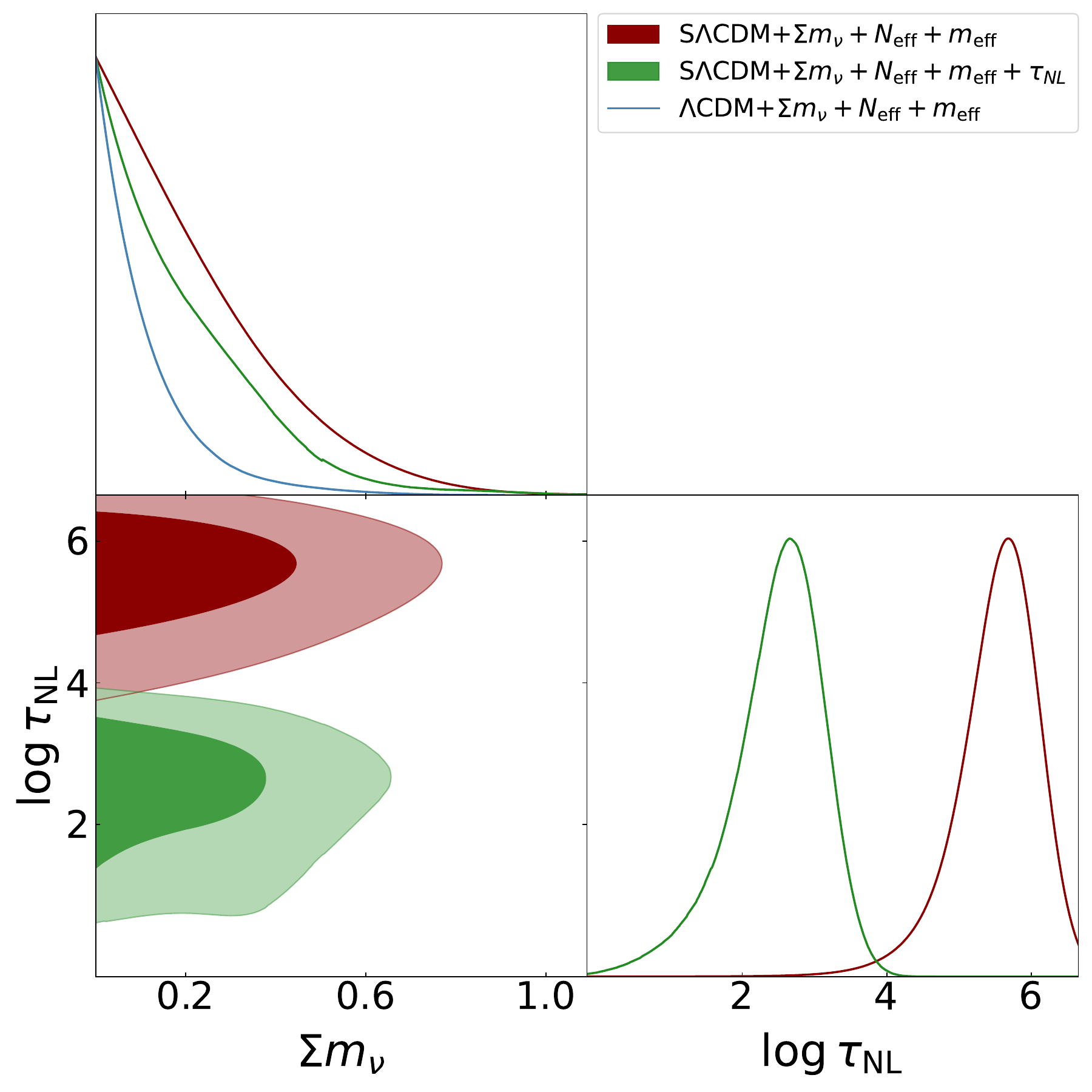}
        \caption{$\Sigma m_\nu+N_{\rm eff}+m_{\rm eff}$}
        \label{fig:3}
    \end{subfigure}
    \caption{Constraints on $\Sigma m_\nu$ for the Planck dataset. Comparing the $\Lambda$CDM case with Super$\Lambda$CDM (S$\Lambda$CDM), including ($+\tau_{NL}$) and not including the prior on the local trispectrum~\cite{Marzouk:2022utf}, we can see that, despite being tighter (from left to right, $~8\%$,  $~15\%$ and $~18\%$ more constraining) when we apply the trispectrum bound, the constraints on the sum of neutrino mass are still more relaxed than the case for $A_0=0$.}
    \label{fig:mnu_tauprior}
\end{figure*}

\subsubsection{$\tau_{NL}$}

\begin{table*}[htp]
    \centering
    \begin{tabular}{lccc}
        \toprule
        Model & Super-$\Lambda$CDM & Super-$\Lambda$CDM$+\tau_{\rm NL}$ & $\Lambda$CDM \\
        \midrule
        $\Sigma m_\nu$ & $< 0.617$\,eV & $< 0.569$\,eV & $< 0.257$\,eV \\
        $\Sigma m_\nu + N_{\rm eff}$ & $< 0.627$\,eV & $< 0.539$\,eV & $< 0.305$\,eV \\
        $\Sigma m_\nu + N_{\rm eff} + m_{\rm eff}$ & $< 0.589$\,eV & $< 0.478$\,eV & $< 0.352$\,eV \\
        \bottomrule
    \end{tabular}
        \caption{Constraints on $\Sigma m_\nu$ [eV] at \ncl for $\Lambda$CDM, \slcdm with the $\tau_{\rm NL}$ prior ($+\tau_{\rm NL}$) and without. We can see that the constraints of neutrino mass, when the similarity is assumed to be nearly one, are still more relaxed than the case where we neglect NG. These values are obtained using Planck dataset alone.}
        \label{tab:mnu_tau}

\end{table*}

In this section, we study how our findings translate into constraints for $\tau_{\rm NL}$ and how the results change when we impose a conservative upper limit on the trispectrum. The trispectrum's shape for quasi-single-field inflation scenario occupies the \textit{intermediate}-shape ground, primarily due to its dependence on $\epsilon$. Our analysis establishes only a lower bound on $\epsilon$, resulting in relatively weak constraints on the amplitude of $\tau_{\rm NL}$. Specifically, using only the Planck likelihood, at \ncl, we find $log{\tau_{\rm NL}}=5.4^{+1.3}_{-1.5}$ for \slcdm and $log \tau_{\rm NL}=5.5^{+1.2}_{-1.4}$ for \slcdm+$m_{\rm eff}$+$\Sigma m_\nu$+$ N_{\rm eff}$. These results are illustrated in \fig{taunl}. 

Previous studies~\cite{Marzouk:2022utf} have established a constraint on $\tau_{\mathrm{NL}}$, setting a limit of $\tau_{NL}<1700$ \ncl, which is 2 orders of magnitude smaller than our central value but consistent with models inside our \ncl contour. To check if these bounds apply to our model, we compute the similarity between the trispectrum shape for our quasi-single.field inflation (QSFI) in \eq{trispectrum} and the standard template (ST)
\begin{equation}
    \vec{T}_{\rm ST}(k_1,k_2,k_3,k_4)=\tau_{\rm NL}P_\zeta(k_1)P_\zeta(k_2)P_\zeta(k_3)P_\zeta(k_4)
\end{equation}
The trispectra are calculated using a grid of $k\in[10^{-4};1]$ and fixing the primordial parameters. Specifically, the similarity corresponds to the value of the cosine:
\begin{equation}
    \text{Similarity}(\epsilon) = \frac{\vec{T}_{\rm QSFI} \cdot \vec{T}_{\rm ST}}{\|\vec{T}_{\rm QSFI}\| \|\vec{T}_{\rm ST}\|}\,.
\end{equation}
It strongly depends on $\epsilon$, as shown in \fig{similarity}. For $\epsilon\sim0$, we recover a good alignment with the standard form. However, for values within our lower bounds, we cannot exclude cases of low similarity, where the constraints are not relevant. 

As qualitatively demonstrated in Fig. 1 in~\cite{Adhikari:2019fvb}, and confirmed in \fig{taunl},  when $\epsilon$ approaches zero (allowing the parameter $a$, as defined in \eq{CSWL}, to approach nearly zero), the current constraint on $\tau_{\mathrm{NL}}$ can still produce a variance consistent with our results. Additionally, the triangular plots show that the constraints on the sum of neutrino masses are less stringent for smaller $\epsilon$. Therefore, it is interesting to examine what happens to our conclusion on the possible bias on $\Sigma m_\nu$ when tight constraints on $\tau_{\rm NL}$ are applied.

Motivated by this consideration, we assumed a similarity of approximately $1$, indicating a nearly perfect alignment with the local shape, making the constraints relevant. Imposing, therefore, an external prior on $\tau_{\rm NL}<1700$ at \ncl we have reevaluated our constraints. Studying all cases where $\Sigma m_\nu$ is a free parameter, we see that although the upper bound for $\Sigma m_\nu$ is lower, it is still not as stringent as in the scenario where NG are not considered. For example, if we let only $\Sigma m_\nu$ vary, the bound  $\Sigma m_\nu<0.617$~eV shifts to $\Sigma m_\nu<0.569$~eV as we apply the constraints on the trispectrum. These results can be seen in \fig{mnu_tauprior} and are listed in \tab{mnu_tau}. We can conclude that the intriguing possibility of a bias in constraining the total neutrino mass persists even when we assume a tight prior on the trispectrum.

We emphasize that a direct and precise analysis of the trispectrum is beyond the scope of this work. However, the flexibility in the model permits exploration within the bounds of a minimal $\epsilon$ value. Similarly, constraints on the trispectrum's contribution~\cite{Agarwal:2013qta,Green:2023uyz} to scale-dependent bias~\cite{Baumann:2012bc} can be bypassed by exploiting the condition where $a$ approaches zero. Since we have not imposed strict constraints on $\epsilon$, we plan to address this aspect in more detail in future work. Finally, it is important to note that, although existing bounds on nonlocal shapes~\cite{Smith:2015uia} exist, they do not directly apply to our specific model framework.

\section{Conclusions} \label{concl}

From the detection of the CMB   to the late-time cosmic acceleration, modern cosmology has witnessed a massive revolution in the understanding of its dynamical evolution. 
The detection of CMB is undoubtedly one of the pioneering discoveries in modern cosmology having great impact on the late-time physics of the Universe. According to the observational evidences, CMB temperature fluctuations follow almost Gaussian pattern as measured by the power spectrum of the density 
fluctuations.   However, deviations from the Gaussian structure of the density fluctuations in the CMB spectrum, known as the primordial non-Gaussianities, can be potential probe to trace the origin and formation of structures in the Universe. Additionally, they can have great impact on the late-time physics, specifically on the tensions between the cosmological parameters.

In this work, we have presented an interesting way to study the NG derived from the quasi-single-field inflationary paradigm, which was first presented by Adhikari and Huterer~\cite{Adhikari:2019fvb}. It consists in promoting the $\lcdm$ model to the \slcdm with the introduction of two additional parameters: $\epsilon$, which comes from the inflationary theory, and $A_0$, a noise parameter with zero average that mimics the NG covariance contribution according to the Super-sample signal. With this setup, we  modified the theoretical code CAMB and explored the parameter space allowing the neutrino sector of the Universe.  We have not found any correlation between NG and $\nnu$ or $\mef$, but the indication for a negative correlation of $A_0$ with $\mnu$ is signaled. In particular, as our analysis suggests a slight indication of $A_0<0$, this implies greater leeway for massive neutrinos. We have then assumed that, despite being in the \textit{intermidiate} shape, there is an alignment with the local shape with a similarity$\sim1$. Then, the trispectrum constraint found in literature can be applied. We saw that, albeit being more stringent, we still obtain more relaxed upper bounds with respect to the case when NG are not considered. Setting $A_0=0$ might consequently skew the constraints on the total neutrino mass, making them overly stringent. 

Concerning the Hubble constant, we find that for Super-$\Lambda$CDM, $H_0$ mildly increases for all the datasets compared to the $\Lambda$CDM-based Planck's estimations but not enough to alleviate it. In addition, a similar observation is found when the sterile neutrino sector is added to Super-$\Lambda$CDM, i.e., for the scenarios Super-$\Lambda$CDM$+\nnu+\meff$ and Super-$\Lambda$CDM$+\nnu+\mnu+\meff$, in which $H_0$ is found to mildly increase for some particular datasets. However, on the contrary, in other extended scenarios, $H_0$ assumes similar values to the $\Lambda$CDM-based Planck~\cite{Planck:2018vyg}. 
On the other hand, the bounds on $S_8$ generally increase in all the scenarios explored in this work, and, thus, we do not see any alleviation of the $S_8$ tension in the present neutrino-based Super-$\Lambda$CDM scenarios. What is interesting is that $A_0$ is confirmed to be different from $0$ at \ncl for Planck, Planck+Pantheon, and also for Planck+BAO, whereas the constraints on $A_0$ from Planck+Lensing and Planck+All are compatible with $0$, suggesting a possible correlation with the lensing problem in Planck. Finally, with the same datasets we were able to set only a lower limit to $\epsilon$.

Based on the outcomes of the present article, one can clearly understand that it will be a pity to avoid the NG from the current cosmic picture.  Even though this new cosmic setup and its extensions are not very sound in the context of cosmological tensions, they do offer new bounds on the neutrinos. According to the existing records, Super-$\Lambda$CDM has just landed to the ocean of cosmological models and it has not received any attention within this short period of its introduction. We trust that this model and its extensions should be further investigated with the upcoming cosmological observations. We are on the way to report some more analyses in this direction.

\acknowledgments 
The authors thank the referee for some useful comments that helped us to improve the quality of the manuscript. 
We thank William Giarè for useful discussions. MF and AM thank TASP, iniziativa specifica INFN for financial support.
EDV is supported by a Royal Society Dorothy Hodgkin Research Fellowship.
SP acknowledges the financial support from  the Department of Science and Technology (DST), Govt. of India under the Scheme  ``Fund for Improvement of S\&T Infrastructure (FIST)'' [File No. SR/FST/MS-I/2019/41].
This article is based upon work from COST Action CA21136 Addressing observational tensions in cosmology with systematics and fundamental physics (CosmoVerse) supported by COST (European Cooperation in Science and Technology).
We acknowledge IT Services at The University of Sheffield for the provision of services for High Performance Computing.

\bibliography{Bibliography}
\end{document}